\documentclass[12pt]{amsart}
\usepackage{amsmath,amsfonts,amssymb,amsthm, amscd,euscript}
\usepackage{tikz-cd}
\usepackage[english]{babel}
\usepackage[T2A]{fontenc}
\usepackage[utf8]{inputenc}
\usepackage{verbatim}
\usepackage{hyperref}
\usepackage{graphicx}
\usepackage{tikz}
\usepackage{fullpage}
\usepackage{color}
\usepackage{soul}
\usepackage{stmaryrd}
\usepackage{bm}
\usepackage{ytableau}
\usepackage{hyperref}
\usepackage{todonotes}
\usepackage{accents}
\usepackage{longtable}

\theoremstyle{plain}
\newtheorem{thm}{Theorem}[subsection]
\newtheorem*{thm*}{Theorem}
\newtheorem{cor}[thm]{Corollary}
\newtheorem{prop}[thm]{Proposition}
\newtheorem{lem}[thm]{Lemma}
\newtheorem*{conj*}{Conjecture}

\theoremstyle{definition}
\newtheorem{defn}[thm]{Definition}
\newtheorem*{notat}{Notation}
\newtheorem{exmp}[thm]{Example}

\newtheorem{rmrk}[thm]{Remark}

\numberwithin{equation}{section}

\newcommand{\nc}{\newcommand}
\newcommand{\rnc}{\renewcommand}
\nc{\ta}{\theta}
\nc{\Id}{{\mathop {\rm Id}}}
\nc{\ad}{{\mathop {\rm ad}}}
\nc{\bra}{\langle}
\nc{\ket}{\rangle}
\nc{\pa}{\partial}
\nc{\ld}{\ldots}
\nc{\cd}{\cdots}
\nc{\hk}{\hookrightarrow}
\nc{\T}{\otimes}
\nc{\gr}{\mathrm{gr}}
\nc{\ov}{\overline}

\newcommand{\bin}[2]{\genfrac{(}{)}{0pt}{}{#1}{#2}}

\nc{\tr} {{\mathop {\rm Trop}}}
\nc{\codim} {{\mathop {\rm codim}}}
\nc{\cO}{\mathcal O}
\nc{\msl}{\mathfrak{sl}}
\nc{\mgl}{\mathfrak{gl}}
\rnc{\U}{\mathrm U}
\nc{\V}{\EuScript V}
\nc{\cL}{\mathcal{L}}

\newcommand\la{\lambda}

\newcommand\al{\alpha}

\newcommand\R{\mathbb{R}}

\newcommand{\bC}{{\mathbb C}}

\newcommand{\bN}{{\mathbb N}}
\newcommand{\bP}{{\mathbb P}}
\newcommand{\bT}{{\mathbb T}}
\newcommand{\bR}{{\mathbb R}}

\newcommand{\bF}{{\mathbb F}}
\newcommand{\cA}{{\mathcal A}}
\newcommand{\cM}{{\mathcal M}}

\newcommand{\argmin}{\text{argmin}}

\usepackage{color}

\begin{document}
\title{THE TROPICAL AMPLITUHEDRON}
\date{\today}
     \author[E. Akhmedova]{Evgeniya Akhmedova}
	\address{Department of Mathematics, Weizmann Institute of Science, Israel}
	\email{evakhmedova@gmail.com}
  \author[R. Tessler]{Ran J. Tessler}
	\address{Department of Mathematics, Weizmann Institute of Science, Israel}
	\email{ran.tessler@weizmann.ac.il}

\maketitle

\begin{abstract}The Amplituhedron is a subspace of the Grassmannian that was recently defined by Arkani-Hamed and Trnka \cite{AHT} in their study of scattering amplitudes in planar $\mathcal{N}=4$ super Yang Mills theory, and was the subject of many papers in the last decade. In this work we define a tropical analog of the amplituhedron, and develop techniques to address it. We prove that many of the key properties of the amplituhedron hold also in this simpler, piecewise linear, model.
\end{abstract}

\tableofcontents

\section{Introduction}
The amplituhedron $\mathcal{A}_{n,k,m}$ is the image of the map $Gr_{k,n}^{\geq 0}\xrightarrow{\;\tilde{Z}\;}Gr_{k,(k+m)}.$ It was introduced by Arkani-Hamed and Trnka \cite{AHT} in the context of scattering amplitudes of $\mathcal{N}=4$ Super Yang Mills (SYM).
This paper defines a tropical analog of the amplituhedron, studies its basic properties and proves tropical analogs to some key properties of amplituhedra.  

We first briefly review the definitions of the nonnegative Grassmannian and its tropical version, and the amplituhedron. We then define the tropical amplituhedron and state some of its properties.

\subsection{The positive Grassmannian}
The \emph{Grassmannian} $Gr_{k,n}(\mathbb{F})$ is the variety of $k$-dimensional linear subspaces in $\mathbb{F}^n$. It can be represented as the quotient space $Gr_{k,n} = GL_k(\mathbb{F})\,\backslash\,\text{Mat}^{\ast}_{k\times n}(\mathbb{F})$, of rank $k$ $k \times n$ matrices modulo invertible row operations.
The projective \emph{Pl\"ucker coordinates} of $V\in Gr_{k,n}$ are defined as follows. For a $k \times n$ matrix $C$ representing $V$, every set $I \subseteq \{1,\dots,n\}$ of $k$ columns defines a maximal $k\times k$ minor. Its determinant 
 is the Pl\"ucker coordinate $P_I(C).$ as the determinant of the $k \times k$ minor corresponding to~$I$. The collection of all $\binom{n}{k}$ Pl\"ucker coordinates yields the \emph{Pl\"ucker embedding} of the Grassmannian into the $\left(\tbinom{n}{k}-1\right)$-dimensional projective space. 
 When the field $\mathbb{F}$ is \emph{ordered}, one defines the he \emph{positive Grassmannian} $Gr_{k,n}^{\geq 0}(\mathbb{F})$ is the subset of the Grassmannian with all $P_I >0$. Its closure is the \emph{nonnegative Grassmannian} $Gr_{k,n}(\mathbb{F})$, with all $P_I \geq 0$.

The theory of positivity for algebraic groups, which contains that of the positive Grassmannian, was developed by Lusztig~\cite{Luz}. It was further studied by Rietsch~\cite{Rie1, Rie2}, Marsh and Rietsch \cite{MR}, Fomin and Zelevinsky~\cite{FZ1,FZ,FZ3}, and Postnikov~\cite{Pos, Pos2}. Postnikov developed a beautiful combinatorial theory of the nonnegative Grassmannian and its stratification into positroid cells. An open \emph{positroid cell} $S \subseteq Gr_{k,n}$
comprises the points where a certain subset of the Pl\"ucker coordinates are positive, i.e., all $C$ such that $P_I(C)>0$ for $I \in \mathcal{M}$ and $P_I(C)=0$ for $I \not\in \mathcal{M}$, for some set~$\mathcal{M}\subseteq\binom{[n]}{k},$ which is called a \emph{positroid} These positroids cells are parameterized by $\mathbb{F}_+^d,$ where $d$ is the dimension of the cell, that can be calculated combinatorially. They correspond to various combinatorial objects such as plabic graphs, Grassmannian necklaces, and permutations. The positive Grassmannian bears relations to diverse areas, including high energy physics, cluster algebras, tropical geometry, and integrable systems \cite{AHBL, AHLS, SW, KoW1, KoW2, LPW, SW2}. 
\subsection{The Amplituhedron}
The amplituhedron, a geometric object discovered by Arkani-Hamed and Trnka \cite{AHT} is a subset of the Grassmannian which is defined as follows. 
For three natural numbers $n,k,m$ satisfying $k+m \leq n,$ and a \emph{positive} matrix $Z \in \text{Mat}^{>0}_{n\times(k+m)}(\mathbb{F}),$ that is a matrix all of its  $\binom{n}{k+m}$ maximal minors have positive determinants. The right multiplication by $Z$ induces a well-defined map $\tilde{Z}:Gr_{k,n}^{\geq 0} \to Gr_{k,(k+m)}$.  The \emph{(tree) Amplituhedron} is the image $Gr_{k,n}^{\geq 0}$ under this map: 
\[\mathcal{A}_{n,k,m}(Z) = \{ CZ : C \in Gr^{\geq 0}_{k,n} \} \subset Gr_{k,(k+m)}. \] 
Since many of the properties of the amplituhedron are known or conjectured to be $Z-$independent we sometimes omit $Z$ from the notation, and just write $\mathcal{A}_{n,k,m}.$
We now list a few well known properties of the amplituhedron.
\begin{itemize}
\item The amplituhedron is a closure of a connected open subspace of $Gr_{k,k+m}$ \cite{GL}.
\item There is a natural collection of $\binom{n}{m}$ functions, the \emph{(momentum) twistor coordinates} roughly defined as follows (see all Definition \ref{def:twistors}): For $Y\in\mathcal{A}_{n,k,m}$ the twistor 
$\bra Y i_1 i_2 \dots i_m \ket$ is the determinant of a matrix with rows $Y_1, Y_2, \dots , Y_k, Z_{i_1}, \dots, Z_{i_m}$ is the determinant of the square $(k+m)\times(k+m)$ matrix obtained by adding the rows $Z_{i_1},\ldots, Z_{i_m}$ to a matrix representation of $Y.$ These coordinates play for the amplituhedron the same role the Pl\"ucker coordinates play for the Grassmannian, see \cite{AHT}.
\item The intersections of the twistors of the form $\bra Y i_1 (i_1+1) i_2(i_2+1) \dots i_{\frac{m}{2}}(i_{\frac{m}{2}+1}) \ket$ are non negative on the amplituhedron, and may vanish only at its boundaries, see \cite{AHT}.
\item For every $Y\in\mathcal{A}_{n,k,m}$ the sequence \[(\bra Y12...(m-1)i\ket)_{i=m}^n\] has exactly $k$ \emph{sign flips} (see \cite{AHTT} and \cite{KW} which extends the definition of sign flips to the case when some twistors in the sequence vanish).
\item The motivation for defining the amplituhedron was to provide a geometric way to calculate $\mathcal{N}=4$ SYM scattering amplitudes. In this calculation \emph{triangulations} of the $m=4$ amplituhedron into images of positroid cells of dimension $4k$ are desireable. 
A triangulation of the amplituhedron $\mathcal{A}_{n,k,m}$ is a collection $\mathcal{T}$, of $km$-dimensional open positroid cells of the nonnegative Grassmannian $Gr^{\geq 0}_{k,n}$, which satisfies the following properties for every positive $n\times (k+m)$ matrix $Z.$
\begin{itemize}
\item \emph{Injectivity}: $S \to \tilde{Z}(S)$ is an injective map for every cell $S \in \mathcal{T}$.
\item \emph{Separation}: $\tilde{Z}(S)$ and $\tilde{Z}(S')$ are disjoint for every two cells $S \neq S'$ in $\mathcal{T}$.
\item \emph{Surjectivity}: $\bigcup_{S \in \mathcal{T}} \tilde{Z}(S)$ is an open dense subset of $\mathcal{A}_{n,k,m}(Z)$.
\end{itemize}
\end{itemize}
The papers \cite{BH,PSBW} study triangulations of the $m=2-$amplituhedron. The papers \cite{EZLT,EZLPSBTW} study triangulations of the $m=4$ amplituhedron.
We refer the reader to \cite{AHT, AHBC1, Hod,BH, KW, LPSV, GL, KWZ, LM, MMP, PSBW,EZLT,EZLPSBTW} for further reading regarding the mathematics and physics of the amplituhedron. Generalizations to non-tree level or other physical theories have been considered \cite{AHBC2, AHTT, AHHS, AHHH, Tr,  AHHL}. Its mathematical foundations are being studied under the name \emph{positive geometries}~\cite{AHBL}.
\subsection{The tropical Grassmannian and its positive part}

Let $\bF_{\bR}$ be the field of real Puiseux series, defined in \cite{Mar}. The \emph{valuation}
\[val:\bF_{\bR}\to\R\] takes a series $f$ to $val(f),$ the lower power of $t$ it contains. $val(0)$ is defined to be $\infty.$ $\bF^+_\bR,$ the \emph{positive part of $\bF_{\bR}$} is the subset of $\bF_\bR$ composed of elements $f$ whose lowest degree term has a positive coefficient. By abuse of notations we will sometimes write $val(x)$ for a vector, matrix, or vector space $x.$ In the first two cases we mean applying the valuation entry-by-entry. In the last case we mean performing the valuation on the Pl\"ucker coordinates. 
\begin{defn}
\emph{The tropical projective space} $\mathbb{TP}^n$ is defined as the quotient of $(\mathbb{R}\cup\{+ \infty\})^{n+1}\setminus \{+\infty, \dots, +\infty\}$ by the translation group $\mathbb{R},$ which acts on each coordinate by translation by the same number
\[s.\dot[P_0:\ldots:P_n]=[P_0+s:\ldots:P_n+s],\]and $\infty+s=\infty.$

    \emph{The tropical Grassmannian} defined in \cite{SW} $\tr Gr_{k,n} \subset \mathbb{TP}^{\binom{n}{k}-1}$ is the image of the Grassmannian $Gr_{k,n}(\bF_\bR)$ under the composition of first applying Pl\"ucker embedding into $\mathbb{P}^{\binom{n}{k}-1}$ and then applying the valuation,
    i.e. \[\tr Gr_{k,n} = \{(val(p_I(V))_{I \in \binom{[n]}{k})} \mid V \in Gr_{k,n}(\bF_\bR)\},\]where $[n]=\{1,2,\ldots,n\},$ and $\binom{[n]}{k}$ stands for the set of $k-$element subset of $[n].$  The \emph{tropical non-negative Grassmannian} is \[\tr Gr_{k,n}^{\geq 0} = \{(val(p_I(V))\mid I \in \binom{[n]}{k}) \mid V \in Gr^{\geq 0}_{k,n}(\bF_\bR)\}.\]  
\end{defn}
There exists another definition of the tropical positive Grassmannian as the subset of $\mathbb{TP}^{\binom{n}{k}-1}$ which satisfy certain tropical three term relations:
\begin{align*}
\left[ 
\begin{array}{ll}
	P_{Sac}+P_{Sbd} = P_{Sab}+P_{Scd} \leq P_{Sad}+P_{Sbc},~~\text{or}  \\
	P_{Sac}+P_{Sbd} = P_{Sad}+P_{Sbc} \leq P_{Sab}+P_{Scd},
\end{array}
\right .
\end{align*}
 where $a < b < c < d$, $S \in \binom{[n]}{k-2},$ $Sab$ denotes $S \cup \{a,b\}$ and $P_{I}$ for $I\in\binom{[n]}{k}$ is the corresponding (tropical) coordinate in $\mathbb{TP}^{\binom{n}{k}-1}.$ The two definitions were shown to agree in \cite{AHLS}.

We can now define \emph{tropical positroid cells} in $\tr Gr_{k,n}^{\geq 0}$ as valuation images of the regular positroid cells. This means that for a positroid $\cM \subset \binom{[n]}{k}$ $S_\cM$ is the set of all $V \in \bT\bR^{\binom{n}{k}-1}$ that satisfy the positive tropical 3-term relations and the positroid condition: $\forall I \notin \cM$ we have  $P_I(V) = +\infty$.

\subsection{The tropical amplituhedron}
The aim of this work is to fill the empty space in the diagram \ref{square}
\begin{align}\label{square}
    \begin{CD}
        Gr^{\geq 0}(\bF_\bR) @>\tilde{Z}>> \cA_{n,k,m}(Z)\\
        @V val VV  @V val VV\\
        \tr Gr^{\geq 0}_{n,k} @>\widetilde{val(Z)}?>> ?
    \end{CD}
\end{align}

For a matrix $Z\in\text{Mat}(\bF_\bR)$ whose maximal minors have positive determinants, we define the \emph{Tropical amplituhedron} as the image of the map
\[val:\mathcal{A}_{n,k,m}(Z))\to \tr Gr_{k,k+m},\]which takes the amplituhedron over $\bF_\bR$ to the tropical Grassmannian. This provides the right vertical arrow in \ref{square}.
In truns out that if we require stronger positivity assumptions on $Z,$ such as that all minor determinants are positive, we can also define the vertical arrow in \ref{square} by
\[\tilde{Z}^{trop}: \tr Gr^{\geq 0}_{k,n} \rightarrow \tr Gr_{k,n},\] where \[(\tilde{Z}^{trop}(C))^J = \underset{I \in \binom{[n]}{k}}{\min}(C^I + val(Z)_I^J).\]
Moreover, with these definitions the diagram commutes, and can be defined directly from the valuation of $Z,$ independently of the lift.
\subsubsection{Properties and main results}We prove that many of the properties of the amplituhedron have a tropical counter part.
\begin{itemize}
\item The tropical amplituhedron contains an open set of full dimension $km,$ see Proposition \ref{full-dim}.  
\item The tropical analogs of the twistor coordinates exists, see Section \ref{subsec:trop_twostor}.
\item The tropical analogs of the boundary twistors satisfy a notion of non negativity on the tropical amplituhedron, and may only vanish on the boundary of the the amplituhedron, see Subsection \ref{subsec:bdry}.
\item For every point $Y$ in the tropical amplituhedron the sequence of tropical twistors\[(\tr\bra Y12...(m-1)i\ket)_{i=m}^n\] has exactly $k$ \emph{tropical sign flips}, see Section \ref{subsec:sign_flip} for exact definitions and proof.
\item Finally, for $m=2$ we prove a tropical-amplituhedron analog of Bao-He's result on triangulations of the $m=2$ amplituhedron \cite{BH}, Theorem \ref{thm:trop_kermit}. To do this, we use techniques similar to those developed in \cite{PSBW}.
\end{itemize}
\subsubsection{Possible applications}
The tropical amplituhedron, in addition to being a natural definition and a space of intrinsic interest, could help studying the amplituhedron. The fact that the complicated geometry of the amplituhedron is replaced by a piecewise linear geometry is likely to simplify studying its properties. Since the tropical amplituhedron can be thought of as a limit of the usual amplituhedron, it is likely that at least $Z-$independent properties of the usual amplituhedron (and many properties of the amplituhedron are expected or proved to be $Z-$independent) might be proven for the tropical version, and then somehow pushed to the non tropical version. Lastly, the motivation for defining the amplituhedron was that for $m=4$ its boundary structure encodes the $\mathcal{N}=4$ planar SYM scattering amplitudes. We believe that the simplified tropical version possesses the same property, and hence could help simplifying even more the calculations of amplitudes.
\subsection{Structure of the paper}
Section \ref{prelim}
provides an extended discussion in tropical matrices and Grassmannian. Section \ref{sec:trop_amp} defines the tropical amplituhedron, and its canonical tropical twistor coordinates. 
We discuss different behaviours of the tropical amplituhedron versus those of the usual amplituhedron. We then show that certain key properties of the usual amplituhedron and twistor coordinates have a tropical counterpart. Subsection \ref{subsec:bdry} shows that the the tropical zero locus of the tropical version of the twistors whose zero loci form boundary strata of the usual amplituhedron - form boundary strata for the tropical amplituhedron. Subsection \ref{subsec:sign_flip} proves a tropical version for the important sign flip property of the amplituhedron (see \cite{AHTT}).

Section \ref{sec:triang} discusses triangulations of the tropical amplituhedron. Its main theorem, Theorem \ref{thm:trop_kermit} proves the existence of triangulations of the $m=2$ tropical amplituhedron by tropical positroid cells of dimensions $2k.$
\subsection{Acknowledgement}
The authors wish to thank Nick Early, Michael Oren-Perlstein and Bernd Sturmfels for stimulating discussions related to this work.
E.A. was supported by the Feinberg Graduate School at Weizmann Institute of Science.
R.T. was supported by the ISF grants no.~335/19 and 1729/23.

\section{Preliminaries}\label{prelim}

\subsection{Tropical positive Grassmannian}

\begin{defn}
    Let $\bF$ be an ordered field. The \emph{Grassmannian} $Gr_{k,n}(\mathbb{F})$ is the space of all $k$-dimensional subspaces of $\mathbb{F}^n$. One of the standard ways to parameterize the Grassmannian is by embedding it into $\bP^{\bin{n}{k}-1}$ using the Pl\"ucker embedding. The space $Gr_{k,n}(\mathbb{F})$ can be considered as a quotient of the space of full rank $k \times n$ matrices by the left action of $GL_k(\bF)$. Pl\"ucker embedding maps an element $V$ in $Gr_{k,n}(\mathbb{F})$,  into the vector $(p_I(V))\mid I \in \bin{[n]}{k})$, where $p_I(V)$ is the $[k]\times I$ minor  of some matrix representative of $V$ that is the same for all $I$.
    \emph{The nonnegative Grassmannian} $Gr_{k,n}^{\geq 0}$ is the subset of Grassmannian 
    comprised of elements with nonnegative Pl\"ucker coordinates. \emph{The positive Grassmannian} $Gr_{k,n}^{> 0}$ is the subset of the Grassmannian
    comprised of elements with nonnegative Pl\"ucker coordinates.

    For $\cM \subset \binom{[n]}{k}$ let $S_\cM$ be the set of $V \in Gr_{k,n}^{\geq 0}$, such that for each $I \in \cM$ the Pl\"ucker coordinate $p_I(V)$ is strictly positive and for each $J \notin \cM$ $p_J(V) = 0$. If $S_\cM \neq \varnothing$, then $\cM$ is called a \emph{positroid} and $S_\cM$ is called a \emph{positroid cell}.
\end{defn}

\begin{exmp}
    The smallest example of a Grassmannian is $Gr_{1,n}(\bF)\simeq\mathbb{FP}^{n-1}$, which is just the collection of linear subspaces of $\mathbb{F}^n$ of dimension $1,$ that is, subspaces spanned by a single non zero vector. The nonnegative part $Gr_{1,n}^{\geq 0}(\bF)$ is the set of all such paces spanned by a vector whose entries are all nonnegative. A positroid cell in $Gr_{1,n}^{\geq 0}(\bF)$ is characterized by setting some of the of the spanning vector's entries to be 0.
\end{exmp}

Next we will describe the tropicalization of the Grassmannian and the positive Grassmannian. There are two general ways to think about tropicalization of algebraic varieties: as a valuation image of an algebraic variety over the field of Puiseux series, or as intersections of zero sets of tropical polynomials. The fundamental theorem of tropical geometry \cite[Theorem 3.1.3]{MSt} states that these two descriptions are equivalent, where relation between them is given by \[\overline{val(V(\mathcal{I}))} = V(Trop(\mathcal{I})),\]
where $\mathcal{I}$ is the corresponding ideal, $Trop(\mathcal{I})$ is its tropicalization, and $V$ stands for zero locus (tropical and non tropical) - see below. We start with the former construction, but will use the latter whenever it is more convenient.

Let $\bC\{\{t\}\} = \cup_{n = 1}^{\infty} \bC ((t^{\frac{1}{n}}))$ be the field of Puiseux series over $\bC$, and $\bR\{\{t\}\}=\cup_{n = 1}^{\infty} \bR ((t^{\frac{1}{n}}))$ be the field of Puiseux series with real coefficients. For a Puiseux series $f \in \bC\{\{t\}\},$ the \emph{valuation} of $f,$ $val(f)$ is the lowest power of $t$ in $f$. Any right inverse map of valuation is called a \emph{lift}. A detailed exploration of the tropical algebraic geometry setup can be found in the textbook \cite{MSt}.

\begin{defn}
    \emph{The tropical projective space} $\mathbb{TP}^n$ is defined as the quotient of the space $(\mathbb{R}\cup\{+ \infty\})^{n+1}\setminus \{+\infty, \dots, +\infty\}$ by the translation group $\mathbb{R},$ which acts on each coordinate by translation by the same number
\[S.\dot[P_0:\ldots:P_n]=[P_0+S:\ldots:P_n+S],\]and $\infty+S=\infty.$

Endow $\bR\cup\{+\infty\}$ with the following topology: the open sets are either open sets in $\bR,$ and sets of the form \[\{x>M\}\cup\{+\infty\},\] for $M\in\bR.$ This induces topology on $(\mathbb{R}\cup\{+ \infty\})^{n+1}$ (the product topology), and  thus also on $(\mathbb{R}\cup\{+ \infty\})^{n+1}\setminus \{+\infty, \dots, +\infty\}.$ We endow $\mathbb{TP}^n$ with the quotient topology.

Observe that the valuation induces a map, also denoted by $val,$ from $\bF_\bR\mathbb{P}^n\to\mathbb{TP}^n.$ This map is well defined since 
\[val([s(t)p_0(t):\dots: s(t)p_n(t)]) = [val(p_0) + val(s): \dots :val(p_n) + val(s)].\]
We define the \emph{charts} $P_i = 0$ of $\bT\bP^n$ as set on points where the $i$-th coordinate is not equal to infinity, and, therefore, can be set to $0$ by adding a constant. These charts are diffeomorphic to $\bR^n$.
\end{defn}

Let $A$ be a finite subset in $\bN^n$ and $f = \sum_{a \in A} c_a x^a$ be a polynomial in $\bR[x_1,\dots, x_n]$, where $x^a = \prod x_i^{a_i}$. A piece-wise linear function obtained from a polynomial $f$ by replacing multiplication with addition and addition with taking a minimum is called a \emph{tropical polynomial} and denoted by $\tr f$.
A \emph{tropical hypersurface} in $(\bR \cup \{+\infty\})^n$ corresponding to $f$, or its \emph{tropical zero locus}, denoted by $V(\tr \ f)$, is the set of points $(x_1, \dots, x_n)$, such that the minimum of $a_1 x_1 + \dots + a_n x_n$ is achieved in at least two values of $a \in A$. For an ideal $\mathcal{I} \subset \bR[x_1,\dots, x_n]$ we denote
\[V(\tr \ \mathcal{I}) = \bigcap_{f \in \mathcal{I}}V(\tr \ f).\]
\emph{The positive part} $\tr V(f)^{\geq 0}$ of a tropical hypersurface $\tr V(f)$ is the set of points in it where minimum of $a_1 x_1 + \dots + a_n x_n$ is achieved both at some $a$  with positive $c_a$ and some $a'$ with negative $c_{a'}$.
\begin{exmp}
    Consider a polynomial $f = x^3 + xy^2 - z$. The corresponding tropical hypersurface is the set \[V( \tr f) = \left\{(x,y,z)| \begin{array}{lll}
         &  3x = x+2y \leq z, \text{ or }\\
         & x+2y = z \leq 3x, \text{ or } \\
         & 3x = z \leq x+2y  
    \end{array} \right\}.\]
    The positive part of this tropical hypersurface is the part of $\tr V(f)$, where $\min(3x,x+2y,z)$ is obtained in $z$ and either $3x$ or $x+ 2y$:

    \[ V(\tr f )^{\geq 0} = \left\{(x,y,z)| \begin{array}{ll}
         & x+2y = z \leq 3x, \text{ or } \\
         & 3x = z \leq x+2y  
    \end{array} \right\}.\]
\end{exmp}
\begin{notat}
    For a polynomial $f \in \bF[x_1, \dots, x_n]$ we denote the set of it's zeroes by $V(f)$. For an ideal $\mathcal{I} \subset \bF[x_1, \dots, x_n]$ we denote the variety corresponding to it as $V(\mathcal{I}) = \cap_{f \in \mathcal{I}}V(f).$
\end{notat}

The following two theorems relate the two constructions of the tropical geometry.
\begin{thm}[Karpanov's theorem]
    For a polynomial $f \in \bC\{\{t\}\}[x_1, \dots, x_n]$ we have 
    \[\overline{val(V(f))} = V(\tr \ f),\]
    where the closure is taken in $\bR^n$ as a Euclidean space.
\end{thm}
\begin{thm}[Fundamental theorem of tropical geometry]
    For an ideal $\mathcal{I} \subset \bC\{\{t\}\}[x_1, \dots, x_n]$ we have 
    \[\overline{val(V(\mathcal{I}))} = V(\tr \ \mathcal{I}),\]
    where the closure is taken in $\bR^n$ as a Euclidean space.
\end{thm}

\begin{defn}\cite{Mar}\label{def:generalised-puis}
    \emph{A generalized Puiseux series} with coefficients from a field $\bF$ is a formal series $\sum_{n=1}^{\infty}a_n t^{b_n},$ where $\forall n~ a_n \in \bF, b_n \in \bR$ and $(b_n)$ diverges to $+ \infty.$
    We will denote the field of generalised Puiseux series with coefficients in $\bC$ as $\bF_\bC$, and with coefficients in $\bR$ as $\bF_\bR$. The field $\bF_\bR$ ordered and \[ 0 < \sum_{n=1}^{\infty}a_n t^{b_n}, ~~~~\text{   if 
     }0 < a_{\underset{n}{\argmin} ~b_n}.\] 
    The valuation map $val: \bF_\bR, \bF_\bC \rightarrow \bR$ is defined the same way it is for Puiseux series:
    \[val(\sum_{n=1}^{\infty}a_n t^{b_n}) = \min_{n}b_n.\]
    The valuation map on these fields induces the norm $||f|| = e^{-val(f)}$. It is shown in \cite[Theorem 11]{Mar} that $\bF_\bR$ and $ \bF_\bC$ are complete with respect to this norm.
    
\end{defn}

It has been shown in \cite{Mar} that if we consider the field of generalized Puiseux series instead of the field of Pisseux series, then the valuation map becomes surjective, meaning

\begin{thm}[Karpanov's theorem for generalized Puiseux series]
    For a polynomial $f \in \bF_\bC[x_1, \dots, x_n]$ we have 
    \[val(V(f)) = V(\tr \ f).\]
\end{thm}
\begin{thm}[Fundamental theorem of tropical geometry for generalized Puiseux series]
    For an ideal $\mathcal{I} \subset \bF_\bC[x_1, \dots, x_n]$ we have 
    \[val(V(\mathcal{I})) = V(\tr \ \mathcal{I}).\]
\end{thm}

From now on we will give definitions for tropical objects as valuations of objects over $\bF_\bR$ or $\bF_\bC$, keeping in mind that most of them were defined as closures of valuation image of objects over $\bR\{\{t\}\}$ or $\bC\{\{t\}\}$ in their originating papers.

\begin{defn}
    \emph{The tropical Grassmannian} $\tr Gr_{k,n} \subset \bT \bP^{\binom{n}{k}-1}$ is the the image of the Grassmannian under the valuation map, applied to the Pl\"ucker coordinates, i.e. \[\tr Gr_{k,n} =  \{(val(p_I(V))_{ I \in \binom{[n]}{k}}) \mid V \in Gr_{k,n}(\bF_\bC)\}.\]
    We view $\tr Gr_{k,n}$ as a topological space, with topology induced from that of $\mathbb{TP}^n$ as a subset. As shown in \cite{SW}, the tropical Grassmannian parameterizes realizable tropical linear spaces, while all tropical linear spaces are parametrized by a much bigger space, the Dressian, the definition of which we remind later in this section. The nonnegative/positive part of this tropical variety is \emph{the nonnegative/positive tropical Grassmannian} \[\tr Gr_{k,n}^{\geq 0/>0} = \{(val(p_I(V))_{I \in \binom{[n]}{k}) }\mid V \in Gr^{\geq 0/>0}_{k,n}(\bF_\bR)\}.\] Let us recall here that the elements of $Gr^{\geq 0}_{k,n}(\bF_\bR) $ are the subspaces for which all Pl\"ucker coordinates have a positive leading coefficient or are equal to zero and $val(0) = +\infty$. 
\end{defn}
\begin{rmrk}
    It is shown in \cite{SW} that, same as in the real case, the dimension of $\tr Gr_{k,n}^{\geq 0}$ is equal to the dimension of $\tr Gr_{k,n}$ and equals $k(n-k)$.
\end{rmrk}

There exists an equivalent definition of the tropical Grassmannian as the intersection of tropical hyperplanes in $(\bT \bP)^{\binom{n}{k}-1}$ corresponding to Pl\"ucker relations.
\begin{defn}
    \emph{A tree-term Pl\"ucker relation} is a Pl\"ucker relation of the type \[p_{Sab}p_{Scd} + p_{Sad}p_{Sbc} = p_{Sac}p_{Sbd},\] where $a < b < c < d$, $S \in \binom{[n]}{k-2}$ and $Sab$ denotes $S \cup \{a,b\}$. \emph{The Dressian} $Dr_{k,n}$ is a space, containing the tropical Grassmannian $\tr Gr_{k,n}$, that is defined as the intersection of all tropical hyperplanes corresponding to three-term Pl\"ucker relations.
    
    Let $P_I$ be coordinates in the tropical space $(\bT \bP)^{\binom{n}{k}-1}$, corresponding to the valuations of $p_I$. The tropicalization of tree-term Pl\"ucker relation $f = p_{Sab}p_{Scd} + p_{Sad}p_{Sbc} - p_{Sac}p_{Sbd}$ is the following system of inequalities, with solutions formed by a union of three half-planes:

\begin{align*}
V(\tr \ f):
\left[ 
\begin{array}{ll}
	P_{Sac}+P_{Sbd} = P_{Sab}+P_{Scd} \leq P_{Sad}+P_{Sbc}, \text{ or}  \\
	P_{Sac}+P_{Sbd} = P_{Sad}+P_{Sbc} \leq P_{Sab}+P_{Scd}, \text{ or}\\
	 P_{Sab}+P_{Scd} = P_{Sad}+P_{Sbc} \leq P_{Sac}+P_{Sbd}.
\end{array}
\right .
\end{align*}
The \emph{positive Dressian} is defined as the intersection of the positive parts of those hyperplanes. 

Since there is only one negative term in three-term Pl\"ucker equations, the positive part of this tropical hyperplane will consist of points satisfying one of the first two inequalities:

\begin{align}
V(\tr \ f) ^{\geq 0}:
\left[ 
\begin{array}{ll}
	P_{Sac}+P_{Sbd} = P_{Sab}+P_{Scd} \leq P_{Sad}+P_{Sbc}, \text{ or}  \\
	P_{Sac}+P_{Sbd} = P_{Sad}+P_{Sbc} \leq P_{Sab}+P_{Scd}.
\end{array}
\right .\label{eq:3_term_positive}
\end{align}
\end{defn}



The dimension of the Dressian is usually much bigger than the dimension of the tropical Grassmannian, however, the same is not true of their positive parts.
\begin{thm}\cite[Theorem 1.3]{AHLS}
    The tropical nonnegative Grassmannian equals the nonnegative Dressian. 
\end{thm}
This means that 3-term Pl\"ucker relations are enough to cut out the nonnegative tropical Grassmannian, but not enough for the whole tropical Grassmannian.

We define \emph{tropical positroid cells} in $\tr Gr_{k,n}^{\geq 0}$ as valuation images of the regular positroid cells. This means that for $\cM \subset \binom{[n]}{k},$ $S_\cM$ is the set of all $V \in \bT\bR^{\binom{n}{k}-1}$ that satisfy the positive tropical 3-term relations and the positroid condition: $\forall I \notin \cM$ we have  $P_I(V) = +\infty$.

\subsection{The Amplituhedron and its triangulations}

The Amplituhedron has been introduced by Nima Arkani-Hamed and Jaroslav Trnka in their 2014 paper \cite{AHT}. 

Let $\bF$ be an arbitrary ordered field. A matrix $Z \in Mat_{n, k}(\bF)$ is called \emph{positive} (\emph{nonnegative}), if each of the \emph{maximal} minors of $Z$ is positive (nonnegative resp.). A matrix $Z \in Mat_{n, k}(\bF)$ is called \emph{totally positive} if all of the minors of $Z$ are positive, and \emph{totally nonnegative} if all of the minors of $Z$ are non-negative. The set of totally positive matrices is denoted by $Mat^{> 0}_{n,k+m}(\bF)$, and the set of totally nonnegative matrices - by $Mat^{\ge 0}_{n,k+m}(\bF)$.

\begin{defn}\cite{AHT}
    
    Let $Z$ be a positive matrix in $Mat^{> 0}_{n,k+m}(\bF)$.
    Consider a map \[\tilde{Z}_{\bF} : Gr^{\geq 0}_{k,n}(\bF) \rightarrow Gr_{k,k+m}(\bF),~~C\mapsto \tilde{Z}([C]) = [CZ],\] 
    such that $
    \tilde{Z}([C]) = [CZ]$, where for $C \in Mat^{\geq 0}_{k,n}(\bF)$ $[C]$ is the point in $Gr^{\geq 0}_{k,n}(\bF)$ represented by it. 
    \emph{The Amplituhedron} $\cA_{n,k,m}=\cA_{n,k,m}(Z)$ is the image of the positive Grassmannian $Gr^{\geq 0}_{k,n}(\bF)$ under the map $\tilde{Z}_{\bF}$.
\end{defn}

\begin{rmrk}
    The fact that $\tilde{Z}_{\bF}([C])$ is, indeed, of dimension $k$ follows from the nonnegativity of $C$ and positivity of $Z$, as shown in \cite{AHT}.
\end{rmrk}

\begin{lem}
If the matrix $Z$ is totally positive, than the amplituhedron $\cA_{n,k,m}(Z)$ is a subset of the nonnegative Grassmannian.
\end{lem}
\begin{proof}
    Let $C \in Mat_{k,n}^{\ge 0}(\bF)$ be a matrix representative for $V\in Gr_{k,n}^{\geq 0},$ and $J \in \binom{[k+m]}{k}.$ By the Cauchy-Binet formula up to multiplication by a scalar, we have $$\tilde{Z}(V)^J = \sum_I C^I Z_I^J.$$
    Since for all $I \in \binom{[n]}{k}$ $C^I \geq 0$ and $Z_I^J > 0$, for each $J$ we have $\tilde{Z}([C])^J \geq 0$, which means that $\tilde{Z}([C])^J \in Gr^{\ge 0}_{k,k+m}(\bF).$
\end{proof}

In addition to the regular Pl\"ucker coordinates, there are other functions which turn out to be the natural coordinates on the Amplituhedron, the so-called \emph{(momentum) twistor coordinates}, defined in \cite{AHT}. They play the same role as the Pl\"ucker coordinates play in the study of the Grassmannian and its positive part. 

\begin{defn}\label{def:twistors}
    Given a matrix $Z \in Mat_{n, k+m}(\bF)$, for a matrix $Y \in Mat_{k,k+m}(\bF)$, the \emph{twistor coordinate} $\bra Y i_1 i_2 \dots i_m \ket$ is the determinant of a $(k+m)\times(k+m)$ matrix with rows $Y_1, Y_2, \dots , Y_k,Z_{i_1}, \dots, Z_{i_m}$. For a vector space $[Y] \in Gr^{\ge 0}_{k,k+m}(\bF)$ the set of twistor coordinates of $Y$ is considered as a vector in projective space $\bF \bP^{\binom{n}{m} - 1}$, which makes it independent of the choice of representing matrix. We will mostly be using twistor coordinates of vector spaces, so it is useful to keep in mind that they are only defined up to multiplication by a scalar.
\end{defn}

\begin{rmrk}\label{rmrk:regular_tw_expansion}
   The twistor coordinates $\tilde{Z}(C)$ can be expressed as follows in terms of minors of $C$ and $Z$ using the Cauchy-Binet formula:
   \begin{align}\label{regular_twistors_expansion}
       \bra \tilde{Z}(C) i_1 i_2 \dots i_m \ket = \sum_{J \in \binom{[n]}{k}} (-1)^{\sigma(J,\{i_1, \dots , i_m\} )} C^J   Z_{J\cup \{i_1, \dots, i_m\}}^{[k+2]},
   \end{align}
   where the summation is over $J = \{j_1 < \dots < j_k\}$ not intersecting $\{i_1,\ldots,i_m\},$ and the sign $(-1)^{\sigma(J\{i_1, \dots , i_m\} )}$ is positive iff the permutation $(j_1j_2\dots j_m i_1\dots i_k)$ is even.
\end{rmrk}

\begin{defn}\label{trian}(\cite{BH})
    Let $\{C_\al\}$ be a collection of positroid cells in $Gr_{k, n}^{\geq 0}$ with \[\dim C_\al = \dim Gr_{k, k+m}=mk.\] We say that $\{C_\al \}$  is a \emph{ triangulation} of the amplituhedron $\cA_{n, k, m}$ if for any initial data $Z \in Mat_{k+m, n}^{>0}$, we have  
\begin{itemize}
\item Injectivity: the map $C_\al \mapsto Z \cdot C_\al$ is injective for any $\al$;

\item Disjointness: $Z \cdot C_\al \cap Z \cdot C_{\al'}=\emptyset$ if $\al \neq \al'$; 

\item Surjectivity: the union of the image $\cup_{\al} Z \cdot C_\al$ is an open dense subset of $\cA_{n, k, m}(Z)$. 
\end{itemize}
\end{defn}

In their paper \cite{BH} Bao and He proved that specific collections of positroid cells, found in \cite{AHTT}, triangulate the $m=2$ amplituhedron. They moreover showed that the cells in each such collection are separated by hyperplanes given by zero loci of certain twistor coordinates.  This was vastly generalized in \cite{PSBW}, where a description of all positroid cells of dimension $2k$ which map injectively into the $m = 2$ amplituhedra was proven, and a large collection of new triangulations was found.

\section{The Tropical Amplituhedron}\label{sec:trop_amp}

In this chapter we review the definition and properties of tropically totally positive matrices, and then define the tropical Amplituhedron. We work out the smallest examples and prove tropical versions of some properties of the regular Amplituhedron. 

\subsection{Tropically totally positive matrices}
In order to define the tropical amplituhedron as the image of a map between tropical spaces, we need to put stricter constraints on the matrix $Z$, namely tropical total positivity. 
\begin{defn}
    A matrix $Z$ in $Mat_{k,n}(\bR \cup \{ \infty \})$ is called \emph{tropically totally positive} if for any lift $\ell: Mat_{k,n}(\bR \cup \{ \infty \}) \rightarrow Mat_{k,n}(\bF_\bR)$ the image $\ell(Z)$ is totally positive in the regular sense, meaning that all minors of $\ell(Z)$ are totally positive.
\end{defn}
The condition of tropical total positivity can be reformulated into surprisingly simple condition on the entries of matrix $Z$.

\begin{notat}
       For a matrix $Z$ we denote by $Z_i^j$ the element in the row $i$ and column $j$.
   \end{notat}

\begin{thm}\label{trop_pos}
    \cite[Lemma 2.11] {GN}
    $Z\in Mat_{k,n}(\bR \cup \{+ \infty \})$ is  tropically totally positive if and only if for any of its minors the minimum of diagonal sums is achieved only on the main diagonal, meaning that $ \forall I,J \subset [n]$ where $|I| = |J|$ and $I = \{i_1 < \dots < i_p\}$, $J = \{j_1 < \dots < j_p\}$ we have
    \begin{align}
       \underset{\sigma \in S_p}{\arg \min} \sum_{t}Z_{i_t}^{j_{\sigma(t)}} = id.\label{eq:trop_pos_gen}
    \end{align}
    Moreover, it is enough to check this condition for all $2 \times 2$ minors:
    \begin{align}
        \forall i < j,\forall a < b,~~Z_i^a + Z_j^b < Z_i^b + Z_j^a.\label{eq:trop_pos_2}
    \end{align}
\end{thm}

\begin{defn}
    For a matrix $Z\in Mat_{n,k+m}(\bR)$, and sets $I = \{i_1 < \dots < i_p\} \subset [n]$ and $J = \{j_1 < \dots < j_p\} \subset [k+m]$ we will write
    \[Z_I^J=\underset{\sigma \in S_p}{\min} \sum_{t}Z_{i_t}^{j_{\sigma(t)}},\] and refer to it as a  \emph{tropical minor} of $Z.$ This is a slight abuse of notation, since for non tropical matrices $\hat{Z}\in Mat_{n,k+m}(\bF_\bR)$ we write $\hat{Z}_I^J$ for its $(I,J)-$minor's determinant. However, this notation is natural since by \cite{GN} for totally positive $\hat{Z}\in Mat_{n,k+m}(\bF_\bR),$ and $Z=val(\hat{Z})$ we have\[val(\hat{Z}^J_I)=Z^J_I,\]
    and we will always know from the context if we consider a matrix in $Mat_{n,k+m}(\bR)$ or in $Mat_{n,k+m}(\bF_\bR).$ 
\end{defn}

We will derive a simple consequence of tropical total positivity in terms of tropical minors.

\begin{lem}
    Suppose $Z \in Mat_{n,n}$ is tropically totally positive and $I, J, L \in \binom{[n]}{k}$, $I \neq J.$ Then
    \begin{align}\label{eq:ineq_for_full_dim}
        Z^I_L + Z^L_J+ Z^J_I > Z_I^I + Z^J_J + Z_L^L.
    \end{align}
\end{lem}

\begin{proof}
    Denote $I = \{i_1 < \dots < j_k\}, J = \{j_1 < \dots < j_k\}$ and $L = \{\ell_1 < \dots < \ell_k\}$.
    By equation \eqref{eq:trop_pos_gen}, we can write out terms of inequality \eqref{eq:ineq_for_full_dim} in terms of elements of matrix $Z$ as follows:
    \begin{align*}
        Z^I_L + Z^L_J+ Z^J_I = \sum_{a \in [k]} Z_{\ell_a}^{i_a} + \sum_{a \in [k]} Z_{j_a}^{\ell_a} +\sum_{a \in [k]} Z_{i_a}^{j_a}, \\
        Z_I^I + Z^J_J + Z_L^L = \sum_{a \in [k]} Z_{i_a}^{i_a} + \sum_{a \in [k]} Z_{j_a}^{j_a} +\sum_{a \in [k]} Z_{\ell_a}^{\ell_a}.
    \end{align*}
    Therefore, 
    \begin{align} \label{eq:full_dim_tech_1}
        (Z^I_L + Z^L_J+ Z^J_I) - (Z_I^I + Z^J_J + Z_L^L) = \sum_{a \in [k]} \left((  Z_{\ell_a}^{i_a} + Z_{j_a}^{\ell_a} +  Z_{i_a}^{j_a}) -   (Z_{i_a}^{i_a} + Z_{j_a}^{j_a} + Z_{\ell_a}^{\ell_a})\right).
    \end{align}

    From inequality \eqref{eq:trop_pos_gen} for any $a \in [k]$ we have 
    \[Z_{\ell_a}^{i_a} + Z_{j_a}^{\ell_a} +  Z_{i_a}^{j_a} > Z_{i_a}^{i_a} + Z_{j_a}^{j_a} + Z_{\ell_a}^{\ell_a}.\]
    Applied to inequality \eqref{eq:full_dim_tech_1}, this yields 
    \[(Z^I_L + Z^L_J+ Z^J_I) - (Z_I^I + Z^J_J + Z_L^L) > 0.\]
\end{proof}
\begin{rmrk}
    Note that the operation of adding a constant to a row or a column of a matrix does not affect its total positivity, since all the diagonal sums in \eqref{eq:trop_pos_gen} are changed by the same constant.
\end{rmrk}

\begin{lem}\label{zeroes_on_diag}
    Any tropically positive matrix $Z \in Mat_{n,k}$ can be transformed to the following form $Z'$ by a sequence of operations of adding a constant to a row or a column.
\begin{align}\label{eq:matrix_type}
    Z' = \begin{pmatrix}
0 & 0 & + & \hdots & +& + \\
+ & 0 & 0 & + &\hdots & + \\
\vdots &  \ddots & \ddots & \ddots & \ddots & \vdots \\
+ & \hdots  & + & 0 & 0 & + \\
+ & \hdots & & + & 0 & 0 \\
+ & & \hdots & & + & 0 \\
* & & & \hdots & & * \\
\vdots & & & & & \vdots \\
* & & & \hdots & & *
\end{pmatrix},
\end{align}

where $+$ signs denote a positive number and $*$ signs denote any real number.
\end{lem}

\begin{proof}
    Let us construct a sequence of tropically totally positive matrices $(Z(s))_{s=1}^{2k}$, where $Z(s+1)$ is obtained from $Z(s)$ by adding a constant to one row or one column,  $Z(1) = Z$ and $Z(2k)$ is of the desired form. We start with making the top left entry zero by subtracting $Z(1)_1^1$ from the first row. Then we obtain $Z(3)$ by subtracting $Z(2)_1^2$ from the second column. Proceeding in this manner, we obtain $Z(2s)$ by subtracting $Z(2s-1)_s^s$ from the $s$-th row of $Z(2s-2)$, and $Z(2s+1)$ by subtracting $Z(2s)_s^{s+1}$ from the $s+1$-th column of $Z(2s)$. Then $Z(2k)$ is the matrix with zeroes on the diagonals as in the form above. We are left with showing that if $Z \in Mat_{n, k}(\bR)$ is a tropically totally positive matrix where for $i \in [k],~Z_i^i = 0 ,$ and for $i\in [k-1],~Z_{i}^{i+1} = 0$, then all other elements in the first $k$ rows of $Z$  are positive.

    To show this, let us point to the condition \eqref{eq:trop_pos_2} of tropical total positivity and note that for any $i<j$, $a < b$ if $Z_i^a, Z_j^b \geq 0$ then at least one of $Z_i^b$ or $Z_j^a$ must be strictly positive. Applying this to our matrix $Z$, since $Z_i^i = 0, Z_{i+1}^{i+1} = 0, Z_{i}^{i+1} = 0$, we have $Z_{i+1}^i > 0$ and since $Z_i^i = 0, Z_{i-1}^{i} = 0, Z_{i}^{i+1} = 0$, we have $Z_{i-1}^{i+1} > 0$. Let us prove the positivity of the rest of the elements by induction.
    Suppose all elements $Z_i^a$ where $ |i - a| < s$ are nonnegative. Then positivity of $Z_i^{i+s}$ follows from $Z_{i}^{i+1} = 0, Z_{i+1}^{i+1} = 0, Z_{i+1}^{i+s} \geq 0$, together with \eqref{eq:trop_pos_2} applied to \[Z_{i}^{i+1} + Z_{i+1}^{i+s} < Z_{i}^{i+s} + Z_{i+1}^{i+1}.\] 
    Similarly, the positivity of $Z_{i+s}^{i}$ follows from \eqref{eq:trop_pos_2}and $Z_{i+s}^{i+s} = Z_{i+s-1}^{i+s} = 0, Z_{i+s-1}^i \geq 0.$ 
\end{proof}

\subsection{The tropical Amplituhedron}

The nonnegative Grassmannian and the Amplituhedron map $\tilde{Z}$ can be defined over any ordered field, so we can consider \[\tilde{Z}: Gr^{\geq 0}_{k,n}(\bF_\bR) \rightarrow \cA_{n,k,m}(Z)\] for $Z \in Mat^{>0}_{n,(k+m)}(\bF_\bR)$. 
\begin{rmrk}
    The proof of the fact that $\dim \tilde{Z}(V) = k$ from \cite{AHT} holds for any ordered field, but it becomes even simpler in the case of totally positive $Z,$ which is the case to which we shall soon restrict:
    since $[C] \in Gr_{k,n}^{\geq 0}(\bF)$ must have at least one non-zero coordinate. Using Cauchy-Binet, the Pl\"ucker coordinates of the image $Y^J = \underset{I \in \binom{[n]}{k}}{\sum} C^I Z_I^J,$ have to be strictly positive.
\end{rmrk}

\begin{defn}
     Let $Z \in Mat^{>0}_{n,(k+m)}(\bF_\bR)$, be a matrix whose valuation $val(Z)$ is tropically totally positive. \emph{The tropical Amplituhedron} $\cA^{trop}_{n,k,m}(Z)$ is the valuation image of $\tilde{Z}(Gr_{k,n}^{\geq 0}(\bF_\bR))$.
\end{defn}
\begin{notat}
     We will denote the tropical Pl\"ucker coordinates on $\tr Gr_{k,k+m}$ containing the image by $Y^I$, and the coordinates on $\tr Gr_{k,n}$ as $C^I$ to avoid confusion between points in $\tr Gr_{k,n}^{\geq 0}$ and $\cA^{trop}_{n,k,m}$.
\end{notat}

\begin{rmrk}
    Since the tropical Amplituhedron is contained in the tropical Grassmannian, we no longer have any matrix representatives for its elements. From now now on we can only think of its points in terms of Pl\"ucker coordinates, however, as we shall see, the matrix elements of $val(Z)$ still play a key role.
\end{rmrk}

\begin{prop}\label{amp_trop_well_defined}
    If $Z$ totally positive, the tropical Amplituhedron $\cA^{trop}_{n,k,m}(Z)$ depends only on $val(Z),$ and not on the whole $Z.$ Moreover, it can be presented as the image of a map \[\tilde{Z}^{trop}: \tr Gr^{\geq 0}_{k,n} \rightarrow \tr Gr_{k,n},\] given by \[(\tilde{Z}^{trop}(C))^J = \underset{I \in \binom{[n]}{k}}{\min}(C^I + val(Z)_I^J).\]
\end{prop}

\begin{proof}
    For any lift $\ell: \tr Gr^{\geq 0}_{k,n} \rightarrow Gr^{\geq 0}_{k,n}(\bF_\bR)$ using Cauchy-Binet formula we can write $\tilde{Z}(\ell(C))^J = \sum_I \ell(C)^I Z_I^J.$ Since $Z$ is totally positive and $\ell(C) \in Gr^{\geq 0}_{k,n}(\bF_\bR)$, we have $\ell(C)^I \geq 0$ and $Z_I^J > 0$. Therefore, 
    \begin{align}
       val(\tilde{Z}(\ell(C))^J) = \underset{I \in \binom{[n]}{k} }{\min}\left(val(\ell(C)^I) + val(Z_I^J)\right)= \underset{I \in \binom{[n]}{k} }{\min}\left(C^I + val(Z_I^J)\right). \label{eq:Pluckers_val}
    \end{align}
    This means that the tropical Amlituhedron can indeed be defined as an image of a map between tropical nonnegative Grassmannians, since $val(\tilde{Z}(\ell(C))^J)$ does not depend on $\ell$.
\end{proof}
\begin{rmrk}
    Note that equation \eqref{eq:Pluckers_val} holds for any $Z$ with positive $k \times k$ minors, which is a weaker condition than tropical total positivity. However, working only with totally positive matrices allows for a more explicit description of the tropical Amplituhedron. For one, tropical total positivity of $val(Z)$ implies $val(Z)^J_I = \sum_{s \in [k]} val(Z^{j_s}_{i_s})$ for $I = \{i_1 < \dots < i_k\} \subset [n]$ and $J = \{j_1 < \dots < j_k\},$  which gives us the expression
    \begin{align*} val(\tilde{Z}(\ell(C))^J) = \underset{I \in \binom{[n]}{k} }{\min}(C^I + val(Z_I^J)) = \underset{I \in \binom{[n]}{k} }{\min}(C^I + \sum_{s \in [k]} val(Z^{j_s}_{i_s})).\end{align*}
\end{rmrk}
We will denote Pl\"ucker coordinates on the image of the tropical amplituhedron map by $Y^I$ for $I \in \binom{[n]}{k}.$

\begin{defn}\label{def:dim} 
     For a point $P$ in the tropical set $S$ we say that $S$ is of \emph{local dimension} $d$ at $P,$ if $d$ is the maximal integer for which every open neighborhood $U\subset S$ of $P$ contains a $d$ dimensional topological manifold. A point $P\in S$ is said to be \emph{regular of local dimension $d$} if the local dimension of $S$ at $P$ is $d,$ and $P$ has a neighborhood in $S$ which is a topological manifold of dimension $d.$ We define the dimension of a set as the maximum of its local dimensions. The sets we are considering in this paper are polyhedral complexes, so their dimension equals the maximum dimension of a polyhedron they contain.    
\end{defn}

\begin{notat}
    For a point $P = [P_0:\dots : P_n] \in \bT\bP^n$, such that $\forall i \in [n]~ P_i \neq +\infty$ we define its \emph{rectangular neighbourhood} $U_\varepsilon(P)$ as the set $\{[Q_0:\dots:Q_n]\mid \forall i~(P_i - P_0)-(Q_i - Q_0) < \varepsilon\}.$
\end{notat}

\begin{prop} \label{full-dim}
    The tropical Amplituhedron $\cA_{n,k,m}(Z)$ is of dimension $km$  for any tropically totally positive matrix $Z \in Mat_{n, k+m}.$
\end{prop}

\begin{proof}
   First, let us show that it is enough to prove the statement for $Z$ of the type $ \forall i \in [k+m]~Z_i^i = Z_i^{i+1} = 0$. By Lemma \ref{zeroes_on_diag}, any totally positive matrix can be transformed into one of this type by adding constants to rows and columns. Adding a constant $a_i$ to the $i$-th column of $Z$ changes the tropical Amplituhedron in the following way: if $Z'$ is the new matrix, then 
   \[\tilde{Z}'(C)^I = \begin{cases}
     \tilde{Z}(C)^I, \text{ if } i \notin I;\\
     \tilde{Z}(C)^I + a_i, \text{ if } i \in I.
   \end{cases}\]
   Since $Z'$ is a tropically totally positive matrix, $Z'(C) \in \tr Gr_{k,k+m}^{\geq 0}.$
   It is straight forward that $\tilde{Z}(C) \mapsto \tilde{Z}'(C)$ is a homeomorphism, hence $\dim \cA^{trop}_{n,k,m}(Z') = \dim \cA^{trop}_{n,k,m}(Z).$
   For $Z''$ being a result of adding a constant $b_i$ to $i$-th row of $Z$, let us define a dual operation on $C \in \tr Gr_{k,n}^{\geq 0}:$
   \[(C'')^I = \begin{cases}
     C^I, \text{ if } i \notin I;\\
     C^I - b_i, \text{ if } i \in I.
   \end{cases}\]
   This operation preserves non-negativity of $C$, since it is equivalent to the composition of a lift from $\tr Gr^{\geq 0}_{k,n}$ to $Gr^{\geq 0}_{k,n}(\bF_\bR)$, multiplication of $i$-th column by a positive constant $t^{-b_i}$ and valuation.
   By definition of $Z''$ and $C''$ for any $I \in \binom{[k+m]}{k}$ we have $\tilde{Z}'' (C'')^I = \tilde{Z}(C),$ so $\dim \cA^{trop}_{n,k,m}(Z'') = \dim \cA^{trop}_{n,k,m}(Z).$

  We first consider the case $n = k+m,$ and postpone the treatment for general $n$ to the end of the proof.
   Fix a totally positive matrix $Z$ of the type $ \forall i \in [k+m],~Z_i^i =0,$ and $\forall i\in[k+m-1],~ Z_{i}^{i+1} = 0.$ 
   Set \[d  = \underset{\substack{i,j \in [k+m] \\ i \neq j-1, j}}{\min}Z_i^j.\]
   For any $I,J \in \binom{[k+m]}{k}$ we either have $Z_I^J = 0$, or $Z_I^J \geq d.$ Note that for all $I$ $Z_I^I = 0$ since all the entries on the main diagonal are zeroes.

   Let us now consider a small neighbourhood of $0 = [0:\dots:0]$ in $\tr Gr_{k,k+m}^{\geq 0}$, denoting $U_\varepsilon(0) \cap \tr Gr^{\geq 0}_{k,k+m}$ by $U^{Gr(k,k+m)}_\varepsilon(0).$ Note that, by definition of a neighbourhood, if $C \in U_\varepsilon^{Gr(k,k+m)}(0)$, then 
\[\forall I, I' \in \binom{[k+m]}{k}~~ |C^I - C^{I'}| < 2\varepsilon.\]
   It is easy to see that the local dimension of $Gr_{k,k+m}^{\geq 0}$ at $0$ coincides with the dimension of $\tr Gr_{k,k+m}^{\geq 0}$. Indeed, for any $C \in \tr Gr_{k,k+m}^{\geq 0},$ and any $\la \in \bR^{>0},$ the scaling of tropical vector space $\la,$ given by \[(\la C)^I = \la C^I\] scales all terms in tropical positivity conditions by $\la$ and thus preserves non-negativity. 
   If $U$ is a bounded open set in the tropical positive Grassmannian, for any rectangular neighborhood $U_\varepsilon$ of $0$ we can find a scaling which sends $U$ to a subset of $U_\varepsilon.$ Since scalings are clearly homeomorphisms, this implies that the local dimension at $0$ is the dimension of $\tr Gr_{k,k+m}^{\geq0},$ equal to $km$.
   

   For a sequence of real numbers $(a) =   (a_1 > a_2 > \dots > a_{m+k})$ let us define an operation $Add_{(a)}$ as follows:
   \begin{align*}
       Add_{(a)}: \tr Gr^{\geq 0}_{k,k+m} &\rightarrow \tr Gr^{\geq 0}_{k,k+m}, \\ \forall C \in \tr Gr^{\geq 0}_{k,k+m}, I \in \binom{[k+m]}{k}:&~~\quad Add_{(a)}(C)^I = C^I + \sum_{i \in I}a_i.
   \end{align*}
   \begin{rmrk}
       The operation $Add_{(a)}$ preserves tropical positivity, since it can be equivalently defined as a composition of a lift, multiplication of $i$-th column by $t^{a_i}$ and then applying the valuation. Since multiplication of a column by a positive scalar preserves regular positivity, the result is again tropically non negative.
   \end{rmrk}
   Let $\varepsilon$ be smaller then $\frac{d}{16k(k+m)}$ and
   small enough that $U^{Gr(k,k+m)}_\varepsilon(0)$ is equidimensional. Fix a sequence $(a) = (a_1, \dots, a_{k+m})$ such that \begin{equation}\label{eq:(a)}\frac{d}{8k} > a_1 > a_2 > \dots > a_{m+k} > 0, \text{ and } \forall i~ a_i - a_{i+1} > 2\varepsilon.\end{equation} Consider the image of $U_\varepsilon(0)$ under $Add_{(a)}.$ If $C \in Add_{(a)}\left( U^{Gr(k,k+m)}_\varepsilon(0) \right)$, then 
   \begin{align}
       \forall I,I' \in \binom{[k+m]}{k}~~ |C^I - C^{I'}| < \frac{d}{4} + 2\varepsilon < \frac{d}{2}.
   \end{align}
   From this inequality follows that if $Z_I^J = 0$ and $Z_{I'}^J > d$, then $C^I + Z_I^J < C^{I'} + Z_{I'}^{J}$. 
   By the special form of $Z,$ we have that $Z_J^J = 0$, for every $J \in \binom{[k+m]}{k}.$ 
   Therefore, for $C \in Add_{(a)} (U^{Gr(k,k+m)}_\varepsilon(0)),$ for all $J \in \binom{[k+m]}{k},$ we have 
   \begin{align} \label{eq:dim_proof_1}
       \text{If } B = \underset{I \in \binom{[k+m]}{k}}{\arg \min}(C^I + Z_I^J), \text{ then } Z_B^J = 0.
   \end{align}
   On the other hand, for any two sets $I \neq H \in \binom{[k+m]}{k}$ and $C \in Add_{(a)}(U^{Gr(k,k+m)}_{\varepsilon}(0))$ we have 
   \begin{align*}
       C^I - C^H \geq \sum_{i\in I}a_i - \sum_{i\in H}a_i - 2\varepsilon.
   \end{align*}
   If $I = \{i_1 < \dots < i_k\}$, $H = \{h_1 < \dots , h_k\}$ and $\forall s~i_s \leq h_s,$ then there must exist $t$ such that $i_t < h_t,$ and thus, using \eqref{eq:(a)},
\begin{align}\label{eq:dim_proof_2}
       C^I - C^H \geq \sum_{i\in I}a_i - \sum_{i\in H}a_i \geq a_{i_t} - a_{h_t} - 2\varepsilon  > 0.
   \end{align}
   Now, for $Z$ of the form of \eqref{eq:matrix_type} if $Z_I^J = 0$ and $I = \{i_1 < \dots < i_k \}, J = \{j_1 < \dots < j_k\}$, then $\forall \ell \in [k]~~Z_{i_k}^{j_k} = 0,$ and hence $i_\ell \in \{j_\ell -1, j_\ell\}.$ Therefore, for any $I$ such that $Z_I^J = 0$ we have $i_\ell \leq j_\ell.$ This, together with \eqref{eq:dim_proof_1} and \eqref{eq:dim_proof_2}  this shows that
   \begin{align*}
      \forall C \in Add_{(a)}\left(U^{Gr(k,k+m)}_\varepsilon(0) \right), \forall J \in \binom{[k+m]}{k}: \quad \underset{I \in \binom{[k+m]}{k}}{\arg \min}(C^I + Z_I^J) = J,
   \end{align*}
   meaning that 
\begin{align}\label{eq:dim_proof_3}
       \forall C \in Add_{(a)}\left(U^{Gr}_\varepsilon(0) \right), \forall J \in \binom{[k+m]}{k}: \quad \tilde{Z}(C)^J = C^J + Z_J^J = C^J.
   \end{align}
    Equation \eqref{eq:dim_proof_3} shows that the map $\tilde{Z}$ on $Add_{(a)}(U^{Gr(k,k+m)}_\varepsilon(0))$ is the identity map, and \[\dim \cA_{k+m,k,m}^{trop}(Z) = \dim Add_{(a)}(U^{Gr(k,k+m)}_\varepsilon(0)) = \dim U^{Gr(k,k+m)}_\varepsilon(0) = km.\]

    In case $n > k+m$, instead of $U^{Gr(k,k+m)}_\varepsilon(0)$, let us consider the set $U_\varepsilon$ of points $C$ in $\tr Gr_{n,k}^{\geq 0}$ which are constructed from points $\hat{C}$ in $U_\varepsilon(0)^{Gr(k,k+m)}$ by setting $C^I = \hat{C}^I$ for $I \subset [k+m]$ and $C^I = +\infty$ otherwise:  
    \[U_\varepsilon = \{[(C^I)_{I \in \binom{[k+m]}{k}}:+\infty:\dots:+\infty]\mid      ~C = (C^I)_{I \in \binom{[k+m]}{k}} \in U_\varepsilon^{Gr(k,k+m)}(0)\}\subset \tr Gr^{\geq 0}_{k,n}.\] 
    Note that this set is contained in the valuation image of elements from $Gr_{k,n}^{\geq 0}(\bF_\bR)$ spanned by rows of a matrix where non-zero entries only appear in the first $k+m$ columns.
    The dimension of $U$ equals the dimension of $U_\varepsilon^{Gr(k,k+m)}(0)$, since the projection $pr : [(C^I)_{I \in \binom{[k+m]}{k}}:+\infty:\dots:+\infty] \mapsto [(C^I)_{I \in \binom{[k+m]}{k}}]$ is a homemorphism onto $U_\varepsilon^{Gr(k,k+m)}(0)$. 
    
    Fix a tropically totally positive matrix $Z \in Mat_{n,k+m}(\bR)$ and consider a matrix $Z_{[k+m]}$ comprised only of the first $k+m$ rows of $Z$. Since the minimum $\min_I(C^I + Z_I^J)$ for $C \in U_\varepsilon$ cannot be obtained in $I$ such that $C^I = +\infty$, for a matrix $\hat{C} \in \tr Gr_{k,k+m}^{\geq 0}$ we have 
    \[\tilde{Z}\left([(\hat{C}^I)_{I \in \binom{[k+m]}{k}}:+\infty:\dots:+\infty]\right) = \tilde{Z}_{[k+m]}(\hat{C}),\]
    and hence
    \[\forall \varepsilon ~~\tilde{Z}(U_\varepsilon) = \tilde{Z}_{[k+m]}(U_\varepsilon(0))\]
    Therefore,
    \[\dim \cA_{n,k,m}^{trop} = \dim \cA_{k+m,k,m}^{trop} = km.\]
    
\end{proof} 
Note that it follows from the previous proposition and Definition \ref{def:dim} that the tropical amplituhedron contains regular points of full local dimension $km.$

\subsection{First example: the case of $k = 1$} \label{exmpl}
For the regular Amplituhedron it follows from \cite{St} that $\cA_{n,1,m}(Z)$ is combinatorially equivalent to the cyclic polytope with $n$ vertices in $\bR^m$, meaning that each of its faces has the maximal number of boundary faces for a polytope with that dimension and number of vertices.

\begin{exmp}\label{exmp-real}
    The left sub diagram of Figure \ref{fig:trop-cyclic} below presents $\cA_{4,1,2}(Z)$ for $Z = \begin{pmatrix}
1 & 0 & 0\\
1 & 1 & 1\\
1 & 2 & 4\\
1 & 3 & 9
\end{pmatrix},$ which is a convex hull of 4 points on the moment curve.
\end{exmp}

The tropical Amplituhedron map $\tilde{Z}$ for $k = 1$ similarly maps the tropical projective space $\bT \bP ^{n-1}$ into a tropical polytope, since it maps \[(C^1: C^2: \dots : C^{n-1} : C^n)\mapsto\left(  \underset{i \in [n]}{\min}(C^i + Z_i^1): \dots :  \underset{i \in [n]}{\min}(C^i + Z_i^{m + 1}) \right),\] which is exactly the definition of the tropical convex hull of the points $Z_1, \dots, Z_n$ in $\bT \bP ^m$ corresponding to the rows of $Z$.

\emph{The tropical cyclic polytope} is the tropical convex hull of points $v_1, \dots ,v_n$ on the tropical moment curve, where $v_{ij} = (i - 1)(m + 1 - j)$ for $i \in [n]$ and $j \in [m+1]$. 
Note that any matrix the rows which are coordinate vectors of the tropical cyclic polytope vertices satisfy the tropical total positivity condition \eqref{eq:trop_pos_2}. Moreover, the tropical convex hull of points given by rows of any tropically totally positive matrix is combinatorially equivalent to a tropical cyclic polytope, since inequalities \eqref{eq:trop_pos_2} dictate the position of the vertices in relation to each other. For more details on the structure of tropical cyclic polytopes see \cite[Chapter 4]{BY}.

\begin{exmp}
    Figure \ref{fig:trop-cyclic} below presents in its left part $\cA_{4,1,2}(Z)$ from Example \ref{exmp-real}, and on the right the tropical Amplituhedron $\cA_{4,1,2}^{trop}(Z')$ for $Z' = \begin{pmatrix}
0 & 0 & 0\\
2 & 1 & 0\\
4 & 2 & 0\\
6 & 3 & 0
\end{pmatrix}$ on the right.
\begin{figure}[ht]
\centering
    \includegraphics[width=0.6\textwidth]{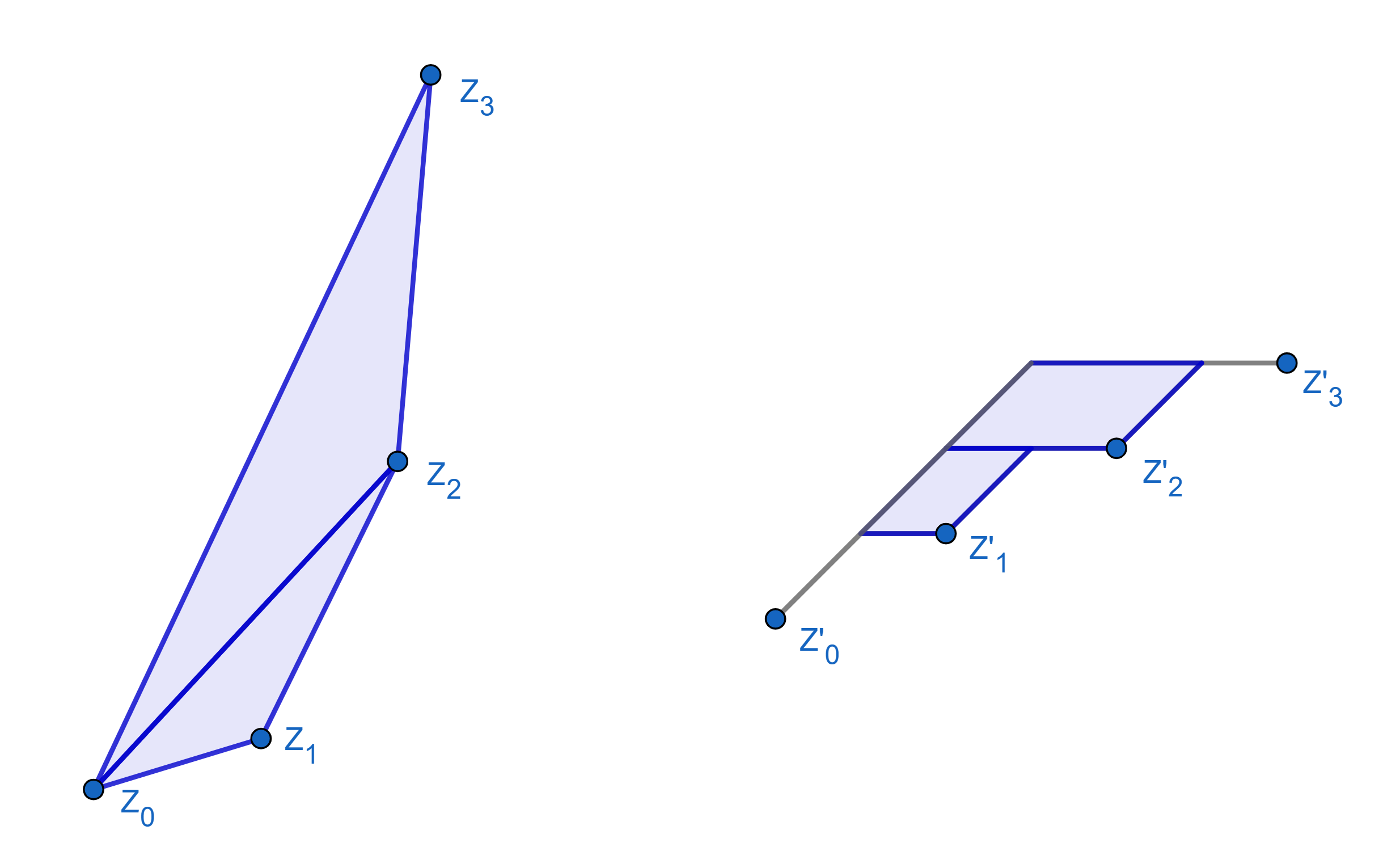}
    \caption{$\cA_{4,1,2}(Z)$ and $\cA^{trop}_{4,1,2}(Z')$}
    \label{fig:trop-cyclic}
\end{figure}
\end{exmp}

A crucial property of $\cA_{n,k,m}$ is that certain positroid cells of $Gr^{\geq 0}_{k,n}$ map injectively under the amplituhedron map and their images triangulate the Amplituhedron. To get a similar picture in the tropical case we have to weaken the injectivity condition. This can be seen even in the simplest case of $k = 1$, and to demonstrate it we will calculate the part of $Gr^{\geq 0}_{1,m}$ that maps injectively into $\cA_{m, 1, m-1}^{trop}$. Since the dimension of the image is equal to the dimension of $Gr^{\geq 0}_{1,m}$, a full-dimension neighbourhood in $Gr^{\geq 0}_{1,m}$ is mapped injectively by the map $\tilde{Z}$ if and only if each coordinate $C^i$ appears as the minimizer in one of the image coordinates. This means that for each $i \in [m]$ there must exist $a \in [m]$ which satisfies 
\[i= \argmin_{j\in [m]}(C^j + Z_j^a).\]
Suppose \[\min_{j \in [m]}(C^j + Z_j^a) = C^i + Z_i^a,\] then for any $t$ we have \begin{equation}\label{eq:i_minimizer_cyclic}C^i + Z_i^a\leq C^t + Z_t^a,\end{equation} this equation is equivalent to \begin{equation}\label{eq:i_minimizer_cycliC^2}Z_i^a - Z_t^a \leq C^t - C^i.\end{equation}  By Proposition \ref{trop_pos} the tropical total positivity of $Z$ gives us inequalities \[Z_i^{b} + Z_t^a \leq Z_i^a + Z_t^{b}\] for each $b > a$ and $t < i$, which implies \[ Z_i^{b} - Z_t^{b} \leq Z_i^a - Z_t^a  \leq C^t - C^i.\] Therefore, using \eqref{eq:i_minimizer_cycliC^2} and the equivalent equation \eqref{eq:i_minimizer_cyclic}, for any $b > a$ the minimum $min(C^j + Z_j^a)$ will be achieved in $j \geq i$. Keeping in mind the injectivity condition, this leaves us with an explicit description of $\tilde{Z}(C)$ for points $C\in \bT\bR^{m-1}$ that have a neigbourhood, on which $\tilde{Z}$ is injective:
\[\tilde{Z} (C^1: \dots : C^m) = (C^1 + Z_1^1 : \dots : C^m + Z_m^m).\] The image of a point $C$ in  $Gr^{\geq 0}_{1,m}$ is of this type if and only if the inequalities 
\begin{align}
    Z_{i+1}^{i+1} - Z_i^{i+1} \leq C^{i} - C^{i+1} \leq Z_{i+1}^i - Z_i^i \label{k=1_inj_condition}
\end{align}
 hold for all $i$. Thus, even in this simplest example, we can see that the structure of areas mapping into full dimension differs between tropical and real cases.
\begin{exmp} The non-injectivity discussed above can be seen even in the smallest expample of an Amplituhedron: while in the regular case $Gr_{1,3}(\bR)^{\geq 0}$ maps into $\cA_{3,1,2}$ injectively, for the tropical nonnegative Grassmanian $\tr Gr_{1,3}^{\geq 0} \simeq \bT \bP^2$ only the area highlited in the figure below maps into full dimension under the amplituhedron map.
\begin{figure}[ht]
\begin{tabular}{ll}
\includegraphics[scale=1.2]{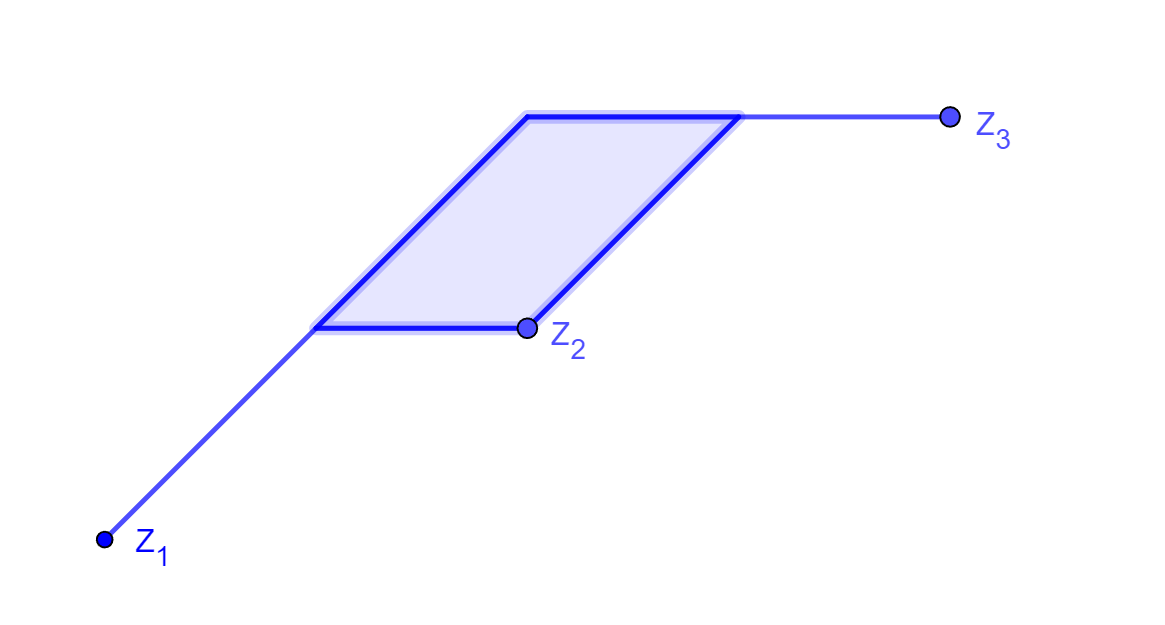}
&
\includegraphics[scale=1.2]{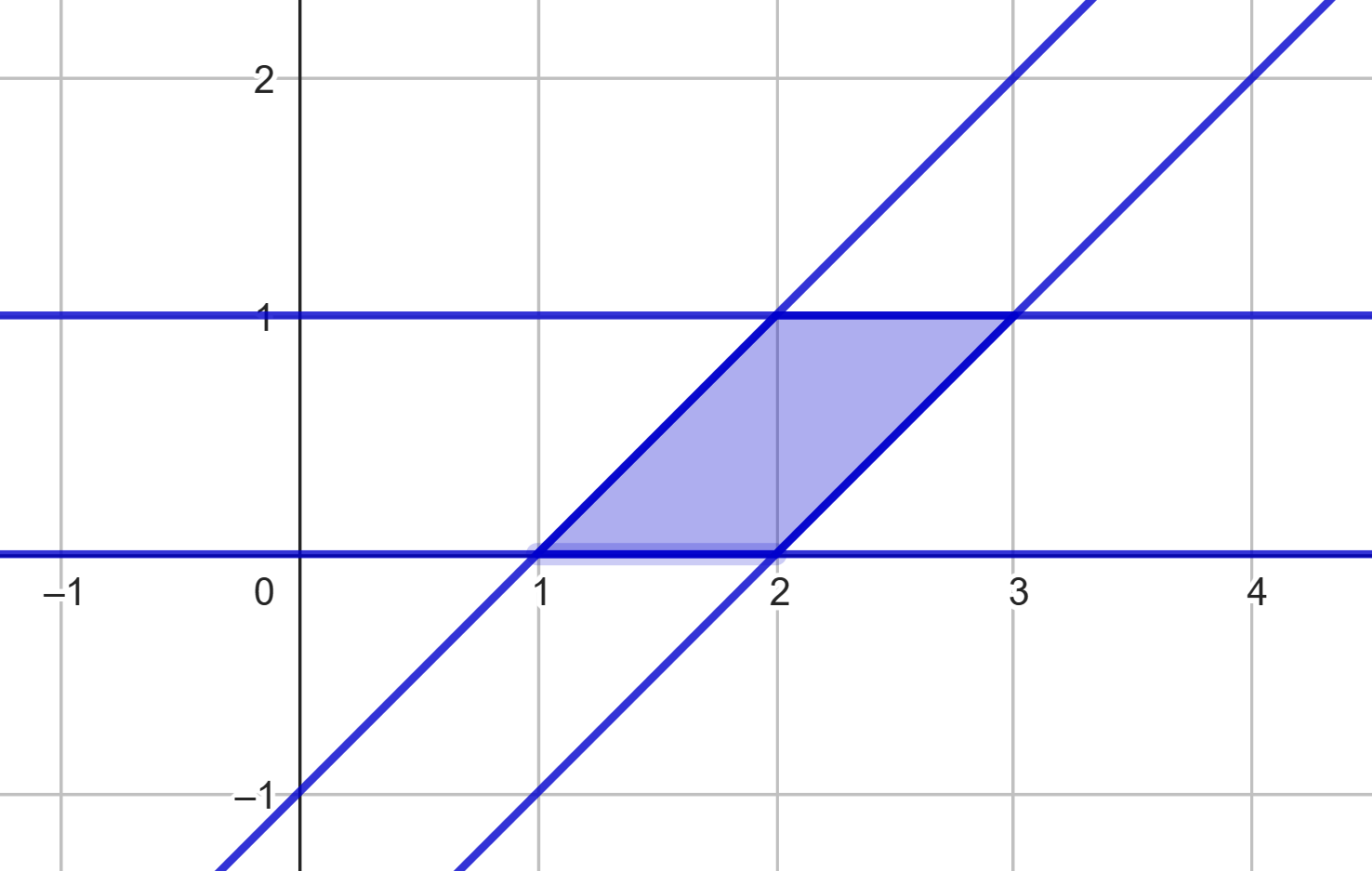}
\end{tabular}
\caption{Left: $\cA_{3,1,2}^{trop}$ in the chart $Y_3 = 0$, \ \ \ 
Right: The region of $\bT \bP^2$ that maps into $\cA_{3,1,2}^{trop}$ injectively in the chart $C^3 = 0$}
\label{Fig:Race}
\end{figure}
\end{exmp}

\subsection{Tropical twistor coordinates}\label{subsec:trop_twostor}

In this section we describe the tropicalization of twistor coordinates from Definition \ref{def:twistors}. In the rest of the section we prove a few combinatorial properties of tropical twistor coordinates, which would be our main instruments in proving statements about the tropical Amplituhedron.

\begin{defn}
    Let $I = \{i_1< \dots < i_m\} \subset [n]$. \emph{The tropical twistor coordinates} are tropical polynomials defined as follows:
\begin{align}
    \tr \bra Y i_1 i_2 \dots i_m \ket = \min_{S \in \binom{[k+m]}{m}} (Y^{[k+m]\setminus S} + Z_I^S). \label{twistor_minimizer}
\end{align}
We call the map taking a point in $\tr Gr^{\geq 0}_{k,n}$ to its set of twistor coordinates in $\bT \bP^{\binom{n}{m}-1}$ the \emph{tropical twistor map}.
\end{defn}
Note that for by the definition of the tropical amplituhedron map $\tilde{Z},$ we have
\[\tr \bra \tilde{Z}(C) i_1 i_2 \dots i_m \ket = \underset{I \in \binom{[n]}{k},S\in \binom{[k+m]}{m}}{\min} (C^I + Z_I^{[k+m]\setminus S} + Z_J^S).\]

\begin{rmrk}\label{trop_twistors=val}
Let $Y_{\bF_\bR} = \tilde{Z}_{\bF_\bR}(C_{\bF_\bR})$, $Y = val(Y_{\bF_\bR})$, $C = val(C_{\bF_\bR})$ and $Z = val(Z_{\bF_\bR})$, and let $J = \{ j_1 < \dots < j_m\}$.
    If $Y \notin  V(\tr \bra Y j_1 \dots j_m \ket)$, then we have 
    \[val(\bra Y_{\bF_\bR} j_1 \dots j_m \ket) =   \tr \bra Y j_1 \dots j_m \ket.\]
\end{rmrk}

\begin{rmrk}\label{fund_thm_for_twistors}
    By the fundamental theorem of tropical geometry, we have 
    \[val(V(\bra Y_{\bF_\bR} j_1 \dots j_m \ket)) =   V(\tr \bra Y j_1 \dots j_m \ket).\]
    Let us recall that $V(\tr \bra Y j_1 \dots j_m \ket)$ is the set of all $Y \in \tr Gr_{k,n},$ for which $\underset{S \in \binom{[n]}{m}}{\arg \min}~(Y^{[k+m]\setminus S}~+~Z_I^S)$ contains at least two sets.
\end{rmrk}

\begin{rmrk}\label{codim_of_twistors}
    In $Gr_{k,k+m}$ an equation of the form $\bra Y j_1 \dots j_m \ket = 0$ cuts out a subvariety of co-dimension $1$. By \cite[Theorem 2.2.3]{EKL} tropicalization of an irreducible variety preserves dimension, so  \[\dim (\tr Gr_{k,k+m} \cap V(\tr \bra Y j_1 \dots j_m \ket)) = km - 1.\] Thus, \[\dim (\cA_{n,k,m}(Z)^{trop} \cap V(\tr \bra Y j_1 \dots j_m \ket)) \leq km-1,\] meaning that by Proposition \ref{full-dim} $ V(\tr \bra Y j_1 \dots j_m \ket)$ intersects $\cA_{n,k,m}^{trop}(Z)$ by a subset of non-zero co-dimension.
    
\end{rmrk}

\begin{notat}
    For sets $I$ and $J$ we will denote by $I \uplus J$ the union of $I$ and $J$ as multisets.
\end{notat}

\begin{lem} \label{diag}
    Fix sets $I \in \binom{[n]}{k}$ and $J \in \binom{[n]}{m}$, such that $I = \{i_1 < \dots < i_k\}$ and $J = \{j_1 < \dots < j_m\}.$ Let $I\uplus J = H = \{h_1 \leq \dots \leq h_{k+m}\}$. Note that no value appears in $H$ more then twice, since all elements in each of $I$ and $J$ are different.
    
    Suppose $$A = \underset{S \in \binom{[k+m]}{m}}{\arg \min}~Z_I^{[k+m] \setminus S} + Z_J^S, \qquad A = \{a_1 < \dots < a_m\}.$$
    Then we must have \[ J = \{h_{a_1}, \dots ,h_{a_m}\}.\]
\end{lem}

\begin{proof}
    Denote $[k+m]\setminus A$ as $Q = \{q_1 < \dots < q_k\}.$
    By equation \eqref{eq:trop_pos_gen} applied to $Z_I^{[k+m] \setminus A}$ and $Z_J^A$, we have
    \[Z_I^{[k+m] \setminus A} + Z_J^A = \min_{\sigma_1 \in S_k}\sum_{t \in [k]} Z_{i_t}^{q_{\sigma_1}(t)} +  \min_{\sigma_2 \in S_m}\sum_{t \in [m]} Z_{j_t}^{a_{\sigma_2}(t)}= \underset{\sigma \in S_{k+m}}{\min}\left(\sum_{t \in [k+m]} Z_{h_{\sigma(t)}}^t\right).\]

    Let $\tau = \underset{\sigma \in S_{k+m}}{\arg \min}\left(\sum_t Z_{h_{\sigma(t)}}^t\right).$
    Suppose there exist $t_1 < t_2$, such that $h_{\tau(t_1)} > h_{\tau(t_2)}$. By \eqref{eq:trop_pos_2} we have \[Z_{h_{\tau(t_1)}}^{t_1} + Z_{h_{\tau(t_2)}}^{t_2} > Z_{h_{\tau(t_2)}}^{t_1} + Z_{h_{\tau(t_1)}}^{t_2}.\] This means that for $\tau' = (t_1 t_2) \circ \tau$ 
    \[\sum_{t\in [k+m]} Z_{h_{\tau(t)}}^t = \sum_{t \in [k+m]} Z_{h_{\tau'(t)}}^t + (Z_{h_{\tau(t_1)}}^{t_1} + Z_{h_{\tau(t_2)}}^{t_2} - Z_{h_{\tau(t_2)}}^{t_1} + Z_{h_{\tau(t_1)}}^{t_2}) > \sum_t Z_{h_{\tau'(t)}}^t,\]
    which contradicts the choice of $\tau$. Thus we have shown that \[h_{\tau(1)} \leq \dots \leq h_{\tau(k+m)},\] therefore, for all $t$ we have $h_{\tau(t)} = h_t$ and $id \in \underset{\sigma \in S_{k+m}}{\arg \min}\left(\sum_{t \in [k+m]} Z_{h_{\sigma(t)}}^t\right).$ Together with \eqref{eq:trop_pos_gen} it gives us 
    \begin{align}
        Z_I^{[k+m] \setminus A} + Z_{j_1}^{a_1} + \dots + Z_{j_m}^{a_m} = Z_I^{[k+m] \setminus A} + Z_J^A = \sum_t Z_{h_{t}}^t, \label{eq:multiset}
    \end{align}
    implying $j_{s} = h_{a_{s}}$ for all $s \in [m]$.
\end{proof}

\begin{notat}
    For multisets $I = \{i_1 \leq \dots \leq i_s\}$ and $J = \{j_1 \leq \dots \leq j_s\}$ and a tropically totally positive matrix $Z$ denote \[\sum_{t = 1}^{s}Z_{i_t}^{j_t} = Z_I^J.\]
    Note that if $I$ and $J$ are sets, then this definition coincides with the tropical minor of $Z$ on rows $I$ and columns $J$ by tropical total positivity condition \ref{eq:trop_pos_gen}, thus there is no abuse of notation.
\end{notat}

\begin{cor}
For $J = \{j_1 < \dots < j_k\} \subset [n]$ we have
    \begin{align}\label{eq:uplus_twistors}
   \tr \bra \tilde{Z}(C) j_1 \dots j_k \ket = \min_{I \in \binom{[n]}{k}}C^I + Z_{I \uplus J}^{[k+m]}.      
    \end{align}
\end{cor}

\begin{prop}\label{not_twistor_zero-no_intersection}
    Let $J = \{j_1 < \dots < j_m\} \subset [n]$ and write \[A = \underset{S \in \binom{[k+m]}{m}}{\arg \min}~Y^{[k+m]\setminus S} + Z_J^S,~~~ B = \underset{I \in \binom{[n]}{k}}{\arg \min}~C^I + Z_I^J,\]
    meaning that 
    \[\tr \bra Y j_1 \dots j_m \ket = C^B + Z_B^{[k+m]\setminus A} + Z_J^A.\]
   Then if $B \cap J \neq \varnothing$ we have $Y \in V(\tr \bra Y j_1 \dots j_m \ket).$
\end{prop}

\begin{proof}
    Denote by $H = \{h_1 \leq \dots \leq h_{k+m}\}$ the union of $B$ and $J$ as multisets. Then by \eqref{eq:multiset} we have 
    \begin{equation}
        C^B + Z_B^{[k+m]\setminus A} + Z_J^A = C^B + \sum_{t \in [k+m]}Z_{h_t}^t.
    \end{equation}
    If $B \cap J \neq \varnothing,$ then $H$ must contain two copies of some number. Let those copies be $h_q = h_{q+1}$. We will treat the case $q \in A, (q+1) \in [k+m]\setminus A.$ The proof for the case $(q+1) \in A, q \in [k+m]\setminus A$ is similar. We have 
    \[\sum_{t \in [k+m]}Z_{h_t}^t = Z_{h_1}^1 + Z_{h_2}^2 + \dots + Z_{h_{q+1}}^q + Z_{h_q}^{q+1} + \dots + Z_{h_{k+m}}^{k+m},\]
    which by Lemma \ref{diag} means that 
    \[C^B + Z_B^{[k+m]\setminus A} + Z_J^A = C^B + Z_B^{[k+m]\setminus (A\setminus{\{q\}} \cup \{q+1\})} + Z_J^{A\setminus{\{q\}} \cup \{q+1\}}.\]
   From equations \eqref{eq:Pluckers_val} and \eqref{twistor_minimizer} we have
    \[ C^B + Z_B^{[k+m]\setminus (A\setminus{\{q\}} \cup \{q+1\})} + Z_J^{A\setminus{\{q\}} \cup \{q+1\}} \leq Y^{[k+m]\setminus (A\setminus{\{q\}} \cup \{q\})} + Z_J^{A\setminus{\{q\}} \cup \{q+1\}} \leq  \tr \bra Y j_1 \dots j_m \ket .\]
    This means that $\underset{S \in \binom{[k+m]}{m}}{\arg \min}~\left(Y^{[k+m]\setminus S} + Z_J^S\right)$ contains both $A$ and $A\setminus{\{q\}} \cup \{q+1\}$ and so $Y$ belongs to $V(\tr \bra Y j_1 \dots j_m \ket).$
\end{proof}

\subsection{Boundary twistors.}\label{subsec:bdry} For the regular amplituhedron Arkani-Hamed and Trnka show that the following inequalities hold for all points of $\cA_{n,k,m}$.
\begin{defn}
    For every $\{i_1 < i_1 + 1 < i_2 < \dots < i_{m/2} < i_{m/2} + 1 \} \in \binom{[n]}{m},$ and $Y\in Gr_{k,k+m}$ the twistor 
    \[\bra Y i_1 (i_1 + 1) \dots i_{m/2} (i_{m/2} + 1) \ket\]is called a \emph{boundary twistor}.
\end{defn}
\begin{prop} \cite{AHT}\label{tw_ineq_original}

    Let $Y \in \cA_{n,k,m}$ and $\{i_1 < i_1 + 1 < i_2 < \dots < i_{m/2} < i_{m/2} + 1 \} \in \binom{[n]}{m}$. Then \[\bra Y i_1 (i_1 + 1) \dots i_{m/2} (i_{m/2} + 1) \ket \geq 0.\]
\end{prop}

In this section we describe a tropical analogue of these boundary twistor inequalities in Proposition \ref{trop_tw_ineq}.
Since the inequalities in \ref{tw_ineq_original} are linear in Pl\"ucker coordinates of $Y$, the tropical version of it will tell us where $\cA_{n,k,m}^{trop}$ lies in relation to a certain tropical hyperplane. Recall that a tropical hyperplane in $\bT \bP ^n$ divides the space into $(n+1)$ disjoint regions, so the twistor inequalities determine which of these regions the tropical Amplituhedron might intersect.

\begin{prop}[\bf{The tropical twistor inequalities}]\label{trop_tw_ineq}  

\hfill

Take $a \in [k+m]$ and $S \in \binom{[k+m]}{m-2},$ such that $\{a-1,a, a+1\} \cap S = \varnothing.$
For every point in $A^{trop}_{n,k,m}(Z)$ and every $i \in [n]$ at least one of the following inequalities holds:
    \begin{align*}
        & Y^{[k+m]\setminus (\{a-1,a+1\}\sqcup S)} - Y^{[k+m]\setminus (\{a,a+1\} \sqcup S)} \geq Z_i^a - Z_i^{a-1}; \\
        & Y^{[k+m]\setminus (\{a-1,a+1\} \sqcup S)} - Y^{[k+m]\setminus (\{a-1,a\} \sqcup S)} \geq Z_{i+1}^a - Z_{i+1}^{a+1}.   
    \end{align*}
\end{prop}
\begin{proof}
Suppose 
\begin{align}\label{eq:pluc_Ran}
    &Y^{[k+2]\setminus (\{a-1,a+1\}\sqcup S)} = C^B + Z_B^{[k+2]\setminus(\{a-1,a+1\}\sqcup S)}.
\end{align}
By \eqref{eq:Pluckers_val}, we have 
\begin{align}
    & Y^{[k+2]\setminus (\{a,a+1\}\sqcup S)} \leq C^B + Z_B^{[k+2]\setminus (\{a,a+1\}\sqcup S)}~~\text{ and } \label{eq:twistor_ineq_pr1}\\
    & Y^{[k+2]\setminus (\{a-1,a\}\sqcup S)} \leq C^B + Z_B^{[k+2]\setminus (\{a,a+1\}\sqcup S)\label{eq:twistor_ineq_pr2}}
\end{align}
Let $Q=[k+m]\setminus(\{a-1,a,a+1\}\sqcup S),$ and write \[Q = \{q_1 < \dots < q_k\},~~B = \{b_1 < \dots b_m\}.\]Let $t$ be the unique index which satisfies $q_t < a < q_{t+1}.$ Equation \eqref{eq:trop_pos_gen}, implies 
\begin{align*}
    Z_B^{Q \sqcup \{a\}}& = Z_{b_1}^{q_1} + \dots + Z_{b_t}^{q_t} + Z_{b_{t+1}}^a + Z_{b_{t+2}}^{q_{t+1}} + \dots + Z_{b_k}^{q_{k-1}}, \\
    Z_B^{Q \sqcup \{a-1\}}& = Z_{b_1}^{q_1} + \dots + Z_{b_t}^{q_t} + Z_{b_{t+1}}^{a-1} + Z_{b_{t+2}}^{q_{t+1}} + \dots + Z_{b_k}^{q_{k-1}}, \\
    Z_B^{Q \sqcup \{a\}}& = Z_{b_1}^{q_1} + \dots + Z_{b_t}^{q_t} + Z_{b_{t+1}}^{a+1} + Z_{b_{t+2}}^{q_{t+1}} + \dots + Z_{b_k}^{q_{k-1}}, \\
\end{align*}
    Subtracting the above equations above one from the other, we obtain
\begin{align*}
    Z_B^{Q \sqcup \{a\}} - Z_B^{Q \sqcup \{a - 1\}} = Z_{b_{t+1}}^a - Z_{b_{t+1}}^{a-1},\\
    Z_B^{Q \sqcup \{a\}} - Z_B^{Q \sqcup \{a+1\}} = Z_{b_{t+1}}^a - Z_{b_{t+1}}^{a+1}.
\end{align*}
Combining these equations with equation \eqref{eq:pluc_Ran} and the inequalities \eqref{eq:twistor_ineq_pr1} and \eqref{eq:twistor_ineq_pr2} yields
\begin{align}
    Y^{[k+m]\setminus (\{a-1,a+1\}\sqcup S)} - Y^{[k+m]\setminus (\{a,a+1\} \sqcup S)} \geq Z_{b_t}^a - Z_{b_t}^{a-1},\label{eq:twist_ineq_pr3} \\
    Y^{[k+m]\setminus (\{a-1,a+1\}\sqcup S)} - Y^{[k+m]\setminus (\{a,a+1\} \sqcup S)} \geq Z_{b_t}^a - Z_{b_t}^{a+1}.\label{eq:twist_ineq_pr4}
\end{align}
Finally, equations \eqref{eq:twist_ineq_pr3} and \eqref{eq:twist_ineq_pr4} together with the condition of tropical total positivity \eqref{eq:trop_pos_2} show
\begin{align*}
    Y^{[k+m]\setminus (\{a-1,a+1\}\sqcup S)} - Y^{[k+m]\setminus (\{a,a+1\} \sqcup S)} \geq Z_{b_t}^a - Z_{b_t}^{a-1} \geq Z_i^a - Z_i^{a-1}, \quad &\text{ if } b_t \leq i; \\
    Y^{[k+m]\setminus (\{a-1,a+1\}\sqcup S)} - Y^{[k+m]\setminus (\{a,a+1\} \sqcup S)} \geq Z_{b_t}^a - Z_{b_t}^{a+1} \geq Z_{i+1}^a - Z_{i+1}^{a+1}, \quad & \text{ if } b_t \geq (i+1).
\end{align*}
Since for each $i \in [n]$ we have either $b_t \leq i$ or $b_t \geq (i+1)$, this proves the proposition. 
\end{proof}

\subsection{Sign-flip condition.}\label{subsec:sign_flip}

For the regular amplituhedron there is a condition on the signs of the sequence $ \{\bra Y 1 2 \dots (m -1) i\ket\}_{i=m}^n.$

\begin{thm} \cite[Section 5]{AHTT}\label{sign-flip-reg}
    For any point $Y \in \cA_{n,k,m}$ the number of sign changes in the sequence $\{\bra Y 1 2 \dots (m -1) i\ket\}_{i=m}^n $ is exactly $k,$ as long as none of the terms is zero.
\end{thm}
\begin{rmrk}
    This condition is generalized in \cite{KW} for the case when the sequence may contain zeroes.
\end{rmrk}

For $m=2,$ this \emph{sign flip constraint} together with the boundary twistors inequalities of Proposition \ref{tw_ineq_original}\footnote{More precisely, for $m=2$ the boundary twistor inequalities are $$\bra Y i (i+1)\ket\geq 0,~i=1,\ldots,n-1$$ and $$(-1)^k\bra Y 1 n\ket\geq 0.$$} were proven to fully describe the $m=2$ amplituhedron (see \cite{AHTT}, \cite{PSBW}). This provides an efficient algorithm for testing whether $Y\in Gr_{k,k+2}$ belongs to $\mathcal{A}_{n,k,2}$. 

We can formulate a similar statement for the tropical Amplituhedron. For that we need to define a tropical analogue of a sign change.

\begin{defn}
$\tr \bra Y 1 2 \dots (m -1) i\ket$ and $\tr \bra Y 1 2 \dots (m -1) (i+1)\ket$ are said to \emph{have the same sign} (or that their sign is the same) if 
\begin{align}
    \underset{S \in \binom{[k+m]}{m}}{\arg \min}~(Y^{[k+m] \setminus S} + Z_{[m-1]\cup \{i\}}^S) = \underset{S \in \binom{[k+m]}{m}}{\arg \min}~(Y^{[k+m] \setminus S} + Z_{[m-1]\cup \{i+1\}}^S), \label{def_sign_flip}.
\end{align}
Otherwise their sign is said to be different.
\end{defn}

\begin{rmrk}
It is easy to see that the minimum in $\underset{S \in \binom{[k+m]}{m}}{\min}~(Y^{[k+m] \setminus S} + Z_{[m-1]\cup \{i\}}^S)$ can be taken over a set much smaller than $\binom{[k+m]}{m}$.
    Note that if 
    \[\tr \bra Y 1 2 \dots (m -1) i\ket = C^B +Z_B^{[k+m] \setminus A}+ Z_{[m-1]\cup \{i\}}^S,\]
    then by Proposition \ref{not_twistor_zero-no_intersection} for $H = B \cup [m-1] \cup \{i\}$ we have $\{h_1,\dots,h_m\} = [m-1]$, which by Lemma \ref{diag} implies that $S = \{1,2, \dots, m-1,a\}$ for some $a > m-1.$ This shows that the following equation is equivalent to equation \eqref{def_sign_flip}:
     \[\underset{m \leq s \leq k+m}{\arg \min}~(Y^{[k+m] \setminus ([m-1] \cup \{s\})} + Z_{[m-1]\cup \{i\}}^{[m-1] \cup \{s\}}) = \underset{m \leq s \leq k+m}{\arg \min}~(Y^{[k+m] \setminus ([m-1] \cup \{s\})} + Z_{[m-1]\cup \{i+1\}}^{[m-1] \cup \{s\}}). \]
\end{rmrk}

In the remainder of this section we will restrict to the case $m = 2$. The general case is treated identically, but requires heavier notation.
 First, let us show that as we go further along the sequence, the index of the Pl\"ucker coordinate of $Y$ minimizing each twistor coordinate increases in lexicographical order.

\begin{lem}\label{lower_bound}
      If \[ \tr \bra Y 1 i \ket = Y^{[k+2] \setminus \{1,a\}} + Z^{\{1,a\}}_{\{ 1,i\}},~\text{and }~  \tr \bra Y 1 (i+1) \ket = Y^{[k+2] \setminus \{1,a'\}} + Z^{\{1,a'\}}_{\{ 1,i +1\}},\] then $a \leq a'$. 
      Thus, there are no more than $k$ sign changes in the sequence $\{\tr \bra Y 1 i\ket\}_{i=2}^n.$
\end{lem}

\begin{proof}
    Since $Y^{[k+2] \setminus \{1,a\}}$ and $Y^{[k+2] \setminus \{1,a'\}}$ minimize respective twistor expressions, we have the following inequalities:
    $$
        \begin{cases}
            Y^{[k+2] \setminus \{1,a\}} + Z^a_i \leq Y^{[k+2] \setminus \{1,a'\}} + Z^{a'}_i \\
            Y^{[k+2] \setminus \{1,a'\}} + Z^{a'}_{i+1} \leq Y^{[k+2] \setminus \{1,a\}} + Z^{a}_{i+1} 
        \end{cases}
    $$
    Combining with the assumptions we see that $Z^a_i + Z^{a'}_{i+1} \leq Z^{a'}_i + Z^{a}_{i+1}.$ Using the tropical total positivity of $Z,$ condition \eqref{eq:trop_pos_2}, this implies $a \leq a'$.
\end{proof}

\begin{lem}\label{upper_bound}
Let $Y \in \cA_{n,k,2}^{trop}$. If for all $I \in [n]$ $Y \notin V(\tr \bra Y 1 i\ket)$, then there are at least $k$ sign changes in the sequence $\{\tr \bra Y 1 i\ket\}_{i=2}^n.$
\end{lem}

\begin{proof}
    Fix a point $C_{\bF_\bR} \in Gr^{\geq 0}_{k,n}(\bF_\bR)$, such that $Y = \tilde{Z}(val(C_{\bF_\bR}))$, and denote $Y_{\bF_\bR} = \Tilde{Z}_{\bF_\bR}(C_{\bF_\bR})$ for some lift $Z_{\bF_\bR}$ of $Z$. Note that since $Y \notin V(\tr \bra Y 1 i\ket)$, we have $\bra Y_{\bF_\bR} 1 i \ket \neq 0$ for all $i$, and, therefore, can apply the sign-flip condition from Theorem \ref{sign-flip-reg}. Pick an $i \in [n]$, such that  $\bra Y_{\bF_\bR} 1 i \ket, \bra Y_{\bF_\bR} 1 (i+1) \ket \neq 0$ and $\bra Y_{\bF_\bR} 1 i \ket$ and $\bra Y_{\bF_\bR} 1 (i+1) \ket$ are of different signs.  As a special case of equation \eqref{regular_twistors_expansion}, when written in terms of $C_{\bF_\bR}$ and $Z_{\bF_\bR}$ the twistor $\bra Y_{\bF_\bR} 1 i \ket$ takes the form
    \[\bra Y_{\bF_\bR} 1 i \ket = \sum_{J \in \binom{[n]}{k}} (-1)^{\sigma(J,i)} C_{\bF_\bR}^J  (Z_{\bF_\bR})_{J\sqcup \{1,i\}}^{[k+2]},\]
    where $\sigma(J,i)$ is defined by the condition that, if $J = \{j_1 < \dots < j_k\},$ then \[j_{\sigma(J,i)} < i < j_{\sigma(J,i)+1}.\]
    Suppose now that there is no sign change between $\tr \bra Y 1 i \ket$ and $\tr \bra Y 1 (i+1) \ket$ meaning that there exists 
    \[B \in \underset{J \in \binom{[n]}{k}}{\arg \min}(C^J + Z_{J \uplus \{1,i\}}^{[k+2]}) \cap  \underset{J \in \binom{[n]}{k}}{\arg \min}(C^J + Z_{J \uplus \{1,(i+1)\}}^{[k+2]}).\]
     By choosing $C$ outside of loci of tropical twistor, we ensured that  \[val(\bra Y_{\bF_\bR} 1 i \ket) = C^B + Z_{B\uplus \{1,i\}}^{[k+2]} \text{ and } val(\bra Y_{\bF_\bR} 1 (i+1) \ket) = C^B + Z_{B\uplus \{1,(i+1)\}}^{[k+2]}.\]
    Since $\bra Y_{\bF_\bR} 1 i \ket$ and $\bra Y_{\bF_\bR} 1 (i+1) \ket$ are of different sign, so must be their minimal degree terms, meaning that $\sigma(B,i) \neq \sigma(B,i+1).$ This, however, yields $ i < b_{\sigma(B,i)} < i+1,$ which is a contradiction.
    Therefore, every sign change in the sequence \[\bra Y_{\bF_\bR} 1 2 \ket,\bra Y_{\bF_\bR} 1 3 \ket, \dots, \bra Y_{\bF_\bR} 1 n \ket\] induces a change in the same spot in the sequence \[\tr \bra Y 1 2 \ket, \tr \bra Y 1 3 \ket, \dots, \tr \bra Y 1 n \ket.\] By Proposition \ref{sign-flip-reg} this proves the lemma.
\end{proof}

\begin{rmrk}\label{step-of-sign-flip}
    Fix $Y \in \cA^{trop}_{n,k,2}$, $Y \notin V(\tr \bra Y 1  i\ket)$ for any $i$.
    Let $(a_i)_{i=2}^n$, $a_i \in \{2, \dots, k+2\}$ be such that $\forall i \ \tr \bra Y 1 i \ket = Y^{[k+2]\setminus \{1,a_i\}}+Z_{\{1,i\}}^{\{1,a_i\}}. $
    By Lemma \ref{lower_bound} the sequence $(a_i)$ is non-decreasing, and by Lemma \ref{upper_bound} it contains no less than $k$ increases. Therefore, $a_1 = 2$, and for each $i$ we have $a_{i+1} \in \{a_i, a_i+1\}.$
\end{rmrk}

From Lemma \ref{lower_bound} and Lemma \ref{upper_bound} immediately follows a tropical version of the sign flip condition.

\begin{thm}\label{sign_flip_2}
    For any point $Y$ in $\cA^{trop}_{n,k,m}$ for which all the tropical twistor coordinates $\tr \bra Y 1 i\ket$ are non-zero, there exactly $k$ sign changes in the sequence 
    \[\tr \bra Y 1 2 \ket , \dots, \tr \bra Y 1 n \ket.\]
\end{thm}

The proofs of Lemmas \ref{lower_bound} and \ref{upper_bound} for general $m$ are identical to the proof of the $m=2$ case, except for replacing $Y^{[k+2] \setminus \{1,a\}}$ by $Y^{[k+m] \setminus \{1, 2, \dots, (m-1), a\}}.$ We  omit the details, but state the conclusion:
\begin{thm}
    For every \[Y\in \cA^{trop}_{n,k,m}\setminus \bigcup_{i\in[n]}V(\tr \bra Y 1 2 \dots (m -1) i\ket),\]the sequence\[ 
    \{\tr \bra Y 1 2 \dots (m -1) i\ket\}_{i=m}^n\]
    has exactly $k$ sign flips.
\end{thm}

\subsection{Injectivity of the tropical twistor map for $m = 2$}
    In the non tropical case the injectivity of the map taking a point in the Amplituhedron to its image in the space of twistors $\mathbb{TP}^{\binom{n}{m}-1}$ is proven in  \cite[Proposition 3.2]{PSBW} by showing that it is a projective linear map given by a matrix of full rank.
    The same reasoning cannot be applied in the tropical Amplituhedron, but the following analog holds.
    \begin{prop}\label{twistors_are_coords}
        Consider $Y, \hat{Y}\in \cA^{trop}_{n,k,2}(Z),$ which are not in the zero locus of any twistor of the type $\tr \bra -i (i+1)\ket$ or $\tr \bra - 1 i \ket$. If for each $i \in [n]$ we have $\tr \bra Y 1 i \ket = \tr \bra \hat{Y} 1 i \ket$ and $\tr \bra Y i (i+1) \ket = \tr \bra \hat{Y} i (i+1) \ket$, then $Y = \hat{Y}$
    \end{prop}
    \begin{proof}
        We will prove injectivity using constructions from \cite{BH,PSBW}.
        \cite{BH} shows that the $m=2$ amplituhedron is triangulated by  a collection of positroid cells (known as the \emph{Kermit triangulation}, considered also in \cite{AHT, AHTT,BH, PSBW,AHBL}. See also Section \ref{subsec:kermit}).
        \cite[Theorem 4.19]{PSBW}   gives an implicit description of the Pl\"ucker coordinates of a point in a Kermit positroid cell, as a polynomial in the twistor coordinates of its image in the amplituhedron. More precisely, for each point $Y$ in the image of one of these Kermit cells \cite[Theorem 4.19]{PSBW}, constructs a preimage \emph{matrix} $C \in \tilde{Z}^{-1}(Y)$ in the Kermit cell, whose entries come from the set of twistors $\pm\bra Y1i\ket,\pm\bra Yi(i+1)\ket.$ Moreover, every twistor from this collection of twistors has a constant sign on the image of the Kermit cell (the sign may depend on which cell we consider). Thus, the Pl\"ucker coordinates $C^I$ are polynomials in $|\bra Y 1 i \ket|$ and $|\bra Y i (i+1) \ket|,$ and they moreover turn out to be with positive coefficients (see \cite[Example 4.20]{PSBW} for an explicit example). 
        These polynomials expressions may fail at the boundaries of the positroid cells, which are shown in \cite{PSBW} to be contained in the zero loci of twistors from the above list. Note that in \cite{PSBW} this is considered over $\bR$, but the proof holds for any ordered field, so we can use this fact over $\bF_\bR$. The absolute value of an element $x$ of $\bF_\bR$ would be simply the nonnegative element of $\{x, -x\}.$


        Let $Y \in \cA^{trop}_{n,k,2}$ be a point in the tropical Amplituhedron outside of zero loci of the tropical twistors of the types $\tr \bra Y 1 i\ket$ and $\tr  \bra Y i (i+1)\ket$. 
        Let $Y_{\bF_\bR} \in \cA_{n,k,2}(\bF_\bR)$ be some lift of $Y.$ By Remark \ref{fund_thm_for_twistors} $Y_{\bF_\bR}$ is outside of zero loci of the corresponding non tropical twistors. Let $C_{\bF_\bR} \in Gr_{k,n}^{\geq 0}(\bF_\bR)$ be the element in the preimage of $Y_{\bF_\bR}$ constructed using the method of \cite{PSBW}.
        Then  for each $I \in \binom{[n]}{k}$ we have an expression (which depends on which Kermit cell contains $C_{\bF_\bR}$).  
        \[C_{\bF_\bR}^I = \sum_{A \in D_I} \prod_{(a_1,a_2) \in A} |\bra Y_{\bF_\bR} a_1 a_2 \ket|\]
        for some $D_I \subset 2^{[n]^2}$, where all elements of $D_I$ are either of the form $(1,i)$ or $(i, i+1)$. The positivity of all summands in these expressions together with Remark \ref{trop_twistors=val} implies 
        \begin{align}
            val(C_{\bF_\bR}^I) = \min_{A \in D_I} \left(\sum_{(a_1,a_2) \in A} val(\bra Y_{\bF_\bR} a_1 a_2 \ket)\right) = \min_{A \in D_I} \left(\sum_{(a_1,a_2) \in A} \tr \bra Y a_1 a_2 \ket \right).\label{eq:twistors_to_pluck}
        \end{align}
        We have such expressions for every point in the image of one of the Kermit cells. Since the images of Kermit cells triangulate the amplituhedron, and their boundaries are contained in the zero loci of twistors of the form $\bra Y_{\bF_\bR}1i\ket,~\bra Y_{\bF_\bR}i(i+1)\ket,$ every amplituhedron point $Y_{\bF_\bR}$ which is not in one of these zero loci has a preimage $C_{\bF_\bR}$ which satisfies \eqref{eq:twistors_to_pluck}. 
        
        Equation \eqref{eq:twistors_to_pluck} allows us to present the valuations of Pl\"ucker coordinates $C_{\bF_\bR}^I$ in terms of the tropical twistors of $Y.$ Equation \eqref{eq:Pluckers_val} thus allows us then to write the tropical Pl\"ucker coordinates of $Y$ in terms of $Z,$ and the collection of twistors $\tr \bra Y 1 i \ket$ and $\tr \bra Y i (i+1) \ket$, proving the proposition.
    \end{proof}
  We can find another analogue of a sign-flip condition for the tropical Amplituhedron, which doesn't have a prototype in the regular case, concerning the sequence of tropical twistor coordinates of the type $\tr \bra Y i (i+1)\ket.$  
  
  \begin{lem}
  If $Y \notin V(\bra Y i (i+1)\ket)$, then the minimizer \[\underset{\{a,b\} \subset [k+2]}{\arg \min}(Y^{[k+2]\setminus \{a,b\} } + Z_{\{i,i+1\}}^{\{a,b\}})\]is of the form $\{a,a+1\}$ for some $a$.
  \end{lem}
  \begin{proof}
      Suppose \[\{a',b'\} \in \underset{\{a,b\} \subset [k+2]}{\arg \min}(Y^{[k+2]\setminus \{a,b\} } + Z_{\{i,i+1\}}^{\{a,b\}})\] and 
      \[Y^{[k+2] \setminus \{a',b'\}} = C^B + Z_B^{[k+2] \setminus \{a',b'\}}.\]
      By Proposition \ref{not_twistor_zero-no_intersection} we have $B \cap \{i,i+1\} = \varnothing.$
      Set $H =B \cup \{i,i+1\}.$ We can write $H = \{h_1 < \dots < h_{k+2}\}$. Then by Lemma \ref{diag} we have $h_{a'} = i, h_{b'} = i+1$. Since $H$ contains every value only once, this means that $b' = a' +1.$

  \end{proof}
\begin{defn}
We say that $\bra Y (i-1) i\ket$ and $\tr \bra Y i (i+1)\ket$ have \emph{the same minimizer} if
\begin{align*}
&\tr \bra Y (i-1) i\ket = Y^{[k+2]\setminus \{a,a+1\} } + Z_{\{i-1,i\}}^{\{a,a+1\}} \text{ and } \\ &\tr \bra Y i (i+1)\ket = Y^{[k+2]\setminus \{a,a+1\} } + Z_{\{i,i+1\}}^{\{a,a+1\}}.
\end{align*}
Otherwise we will say that a minimizer change occurs between $\bra Y (i-1) i\ket$ and $\bra Y i (i+1)\ket$. By analogy with sign-flip condition, we will call this change \emph{min-flip.}

\end{defn}  

\begin{lem}\label{weird_lower_bound}
    If for $a' \neq a$ we have  \begin{align*}
&\tr \bra Y (i-1) i\ket = Y^{[k+2]\setminus \{a,a+1\} } + Z_{\{i-1,i\}}^{\{a,a+1\}} \text{ and } \\ &\tr \bra Y i (i+1)\ket = Y^{[k+2]\setminus \{a',a'+1\} } + Z_{\{i,i+1\}}^{\{a',a'+1\}},
\end{align*}
then $a' > a.$ Thus, there are no more than $k$ min-flips in the sequence $\{\tr \bra Y i (i+1)\ket\}_{i=1}^{n-1}.$
\end{lem}
The proof of this lemma is similar to that of Lemma \ref{lower_bound}, and will be omitted.

\begin{lem}\label{weird_upper_bound}
For any point
\[Y \in \cA_{n,k,2}^{trop}\setminus \left(\bigcup_{i\in[n]}{V(\tr \bra Y 1 i\ket)}\cup\bigcup_{i\in[n]}{V(\tr \bra Y i (i+1)\ket)}\right),\] which is regular of full local dimension $2k$ in $\cA^{trop}_{n,k,2}$, there are at least $k$ min-flips in the sequence $\{\tr \bra Y i (i+1)\ket\}_{i=1}^{n-1}.$
\end{lem}

\begin{proof}
    This fact follows from Proposition \ref{twistors_are_coords}. 
    The twistors $\tr \bra Y 1 i\ket$ only depend on the coordinates of the form $Y^{[k+2] \setminus \{1,a\}},$ while the twistors $\tr \bra Y i (i+1)\ket$ only depend on coordinates of the type $Y^{[k+2] \setminus \{a, a+1\}}.$ The total number of these Pl\"ucker coordinates is $2k+1.$ Since the local dimension of $\cA_{n,k,2}^{trop}$ in $Y$ is $2k$, to determine a point in a neighbourhood of $Y$ uniquely we need at least $2k+1$ Pl\"ucker coordinates, as those coordinates are projective. Therefore, if  for some $a$ the coordinate $Y^{[k+2] \setminus \{a, (a+1)\}}$ doesn't appear as a summand in any of $\tr \bra Y i (i+1)\ket$, then the twistors of types $\tr \bra Y 1 i\ket$ and $\tr \bra Y i (i+1)\ket$ determine at most $2k$ Pl\"ucker coordinates in a neighbourhood of $Y$, which contradicts the injectivity of the twistor map, and the fact that $Y$ is regular of local dimension $2k.$ 
\end{proof}

Lemma \ref{weird_lower_bound} and \ref{weird_upper_bound} combine to yield the following min-flip condition for tropical twistor coordinates of the type $\tr \bra i (i+1)\ket$.
\begin{thm}\label{weird_signflip}
    For any point \[Y\in\cA^{trop}_{n,k,2}\setminus\left(\bigcup_{i}V(\tr \bra Y i (i+1)\ket)\cup\bigcup_i V(\tr \bra Y 1 i \ket)\right),\] in which the local dimension of $\cA^{trop}_{n,k,2}$ is $2k$, the number of min flips in the sequence \hfill
    $\{\tr \bra Y i (i+~1)\ket\}_{i=1}^{n-1}$ is exactly $k$.
\end{thm}

\begin{rmrk}[Similar to Remark \ref{step-of-sign-flip}]\label{step-of-min-flip}

    Fix $Y \in \cA^{trop}_{n,k,2}$, $Y \notin V(\tr \bra Y i  (i+1)\ket)$ for any $i$.
    Let $(a_i)_{i=2}^n$, $a_i \in \{2, \dots, k+2\}$ be such that $\forall i \ \tr \bra Y i (i+1) \ket = Y^{[k+2]\setminus \{a_i,a_i+1\}}+Z_{\{i,i+1\}}^{\{a_i,a_i+1\}}. $
    By Lemma \ref{weird_lower_bound} the sequence $(a_i)$ is non-decreasing, and by Lemma \ref{weird_upper_bound} it contains no less than $k$ increments. Therefore, $a_1 = 2$, and for each $i$ we have $a_{i+1} \in \{a_i, a_i+1\}.$
\end{rmrk}

\begin{notat}
    For $Y\in\cA^{trop}_{n,k,2}\setminus\left(\bigcup_{i}V(\tr \bra Y i (i+1)\ket)\cup\bigcup_i V(\tr \bra Y 1 i \ket)\right),$ 
    denote by $\mathfrak{m}_t=\mathfrak{m}_t(Y,Z)$ the place of the $t$-th minimizer change in the sequence \\ $\{\tr \bra Y i(i+1)\ket\}_{i=2}^n.$ That is, the $t$-th min-flip happens between \\
    $\tr \bra Y (\mathfrak{m}_t -1) \mathfrak{m}_t \ket$ and $\tr \bra Y \ \mathfrak{m}_t (\mathfrak{m}_t +1) \ket.$

    In a similar fashion, we denote by $\mathfrak{s}_t=\mathfrak{s}_t(Y,Z)$  the indicators of the sign-flip, so that the $t$-th sign change in the sequence  $\{\tr \bra Y 1 i\ket\}_{i=2}^n$ happens between $\tr \bra Y 1 (\mathfrak{s}_t-1) \ket $ and $\tr \bra Y 1 \ \mathfrak{s}_t \ket.$
\end{notat}

\begin{rmrk}
    The indicators $\mathfrak{m}_t$ and $\mathfrak{s}_t$ point to the places of minimizer change in $\{\tr \bra Y i(i+1)\ket\}_{i=2}^n$ and $\{\tr \bra Y 1 i\ket\}_{i=2}^n$ respectively, and from Remarks \ref{step-of-min-flip} and  \ref{step-of-sign-flip} we see that it means
    \begin{align*}
        \underset{S \in \binom{[k+2]}{2}}{\arg \min} (Y^{[k+2]\setminus S} + Z_{\mathfrak{m}_t,\mathfrak{m}_t+1}^S )&= \{t+1,t+2\},\\
         \underset{S \in \binom{[k+2]}{2}}{\arg \min} (Y^{[k+2]\setminus S} + Z_{1,\mathfrak{s}_t}^S) &= \{1,t+2\}
    \end{align*}

    Or, equivalently,
    \begin{align*}
        \tr \bra Y \ \mathfrak{m}_t,\mathfrak{m}_t+1 \ket &= Y^{[k+2] \setminus \{t+1,t+2\}} + Z_{\{\mathfrak{m}_t,\mathfrak{m}_t+1\}}^{\{t+1,t+2\}},\\
        \tr \bra Y 1 \ \mathfrak{s}_t \ket &= Y^{[k+2] \setminus \{1,t+2\}} + Z_{\{1,\mathfrak{s}_t\}}^{\{1,t+2\}}.
    \end{align*}

   Therefore, for each $t \in \{2, \dots, k+1 \}$ we have 
   \begin{align}
       Y^{[k+2]\setminus \{t,t+1\}} =  \tr \bra Y \ \mathfrak{m}_{t-1},\mathfrak{m}_{t-1}+1 \ket - Z_{\{\mathfrak{m}_{t-1},\mathfrak{m}_{t-1}+1\}}^{\{t,t+1\}},\label{eq:mf-y's}
   \end{align}
   and for each $t \in \{3, \dots, k+2\}$ we have 
   \begin{align}
       Y^{[k+2] \setminus \{1,t\}} =  \tr \bra Y 1 \ \mathfrak{s}_{t-2} \ket - Z_{\{1,\mathfrak{s}_{t-2}\}}^{\{1,t\}}\label{eq:sf-y's}
   \end{align}
Additionally we have 
   \[Y^{[k+2]\setminus \{1,2\}} = \tr \bra Y 1 2\ket - Z_{\{1,2\}}^{\{1,2\}},\]
   which together with equations \eqref{eq:mf-y's} and \eqref{eq:sf-y's} shows that all tropical  Pl\"ucker coordinates $Y^{[k+2]\setminus \{1,t\}}$ and $Y^{[k+2]\setminus \{t,t+1\}}$ of a point in $\cA_{n,k,2}^{trop}(Z)$ are uniquely determined by the values of twistor coordinates
   \begin{align}
      \{ \tr \bra Y 1 2\ket \} \cup \{\tr \bra Y \ \mathfrak{m}_t,\mathfrak{m}_t+1 \ket | t \in [k]\} \cup \{\tr \bra Y 1 \ \mathfrak{s}_t \ket| t \in [k]\}.
   \end{align}
\end{rmrk}

\section{Triangulations of $\cA_{n,k,2}^{trop}$}\label{sec:triang}
We have already seen in the example in Section \ref{exmpl} that there isn't always a collection of $2k$-dimensional tropicalizations of positroid cells in $\tr Gr_{k,n}^{\geq 0}$ mapping injectively under the tropical amplituhedron map. Therefore, a triangulation of the tropical Amplituhedron by $\tilde{Z}$-images of positroid cells in the sense of Definition \ref{trian} does not exist. However, triangulations do exist if we slightly relax their definition:

\begin{defn}\label{def:trop_triangulation}
    Let $\{S_\al\}_{\alpha\in \mathcal{T}}$ be a collection of cells in $\tr Gr_{k, n}^{\geq 0}$ with \[\dim S_\al = \dim Gr_{k, k+m}=km.\] We say that $\{S_\al \}$ yields a \emph{triangulation} of the tropical amplituhedron $\cA_{n, k, m}^{trop}$ if for any initial data $Z \in Mat_{k+m, n}^{>0}$, we have  
\begin{itemize}
\item (Weak) injectivity: for each $S_\al$ there exists a decomposition, which might be $Z$-dependant, \[S_\al=S_\al^{int} \bigsqcup S_\al^{bound},\] where $S_\al^{int}=S_\al^{int}(Z)$ is open, $\tilde{Z}$ is injective on $S_\al^{int},$ and the $\tilde{Z}-$image of  $S^{bound}_\al = S^{bound}_\al(Z)$ is contained in the union of zero loci of finitely many tropical polynomials which are not identically zero on $\tr Gr_{k,k+m}.$ In particular \[\dim(\tilde{Z}(S^{bound}_\al)) < \dim \cA^{trop}_{n,k,m}(Z).\] 
\item (Weak) disjointness (or separation): $\tilde{Z}(S_\al^{int}) \cap \tilde{Z}( S_{\al'}^{int})=\emptyset$ if $\al \neq \al'.$

\item Surjectivity:  $\cA^{trop}_{n, k, m}(Z)=\overline{\cup_{\al\in\mathcal{T}} \tilde{Z} (S_\al)}$. 
\end{itemize}
\end{defn}


Note that the cells $S_\al^{int}$ form a triangulation in the regular sense of a subset of $\cA^{trop}_{n,k,m}(Z)$ whose complement is of positive codimension. 

To construct positroid triangulations of $\cA^{trop}_{n,k,2}(Z)$ we will be using the Kermit triangulation considered in \cite{ AHT, AHTT,BH, PSBW,AHBL}. These collections of cells were proven to triangulate the $m=2$ amplituhedra in \cite{BH}, and via different methods also in \cite{PSBW}. We should point out that even though the above references considered amplituhedra over $\bR$, the exact same proofs and definitions work for any ordered field, including $\bF_\bR$, a fact we shall use.


In the following subsections we will first describe the construction of a tropical triangulation for $\cA_{n,k,2}^{trop}(Z),$ and then prove (weak) injectivity of $\tilde{Z}$ on each cell, (weak) separation between images of different cells and surjectivity of $\tilde{Z}$ for the collection of cells.

\subsection{The Kermit triangulation}\label{subsec:kermit}
In this subsection we will introduce a parametrization on certain tropical positroid cells, inspired by \cite{PSBW}. 
We start with describing the collection of positroid cells triangulating $\cA_{n,k,2}$. For a detailed explanation of this construction (for more general cells) we refer to \cite{PSBW}.

\begin{defn}
    Let $\mathcal{P}_n$ be a convex $n$-gon with vertices labeled from $1$ to $n$. By $\mathcal{T}_{a,b,c}$ we will denote a \emph{triangle} with vertices $a,b,c$. We will call a collection $\tau$ of $k$ different triangles in $\mathcal{P}_n$ of the type $\mathcal{T}_{1,a,(a+1)}$ a \emph{$(k,n)$ - tiling}. For each $i \in [n] \setminus\{1\}$ the number of triangles in $\tau$ to the left of the diagonal $(1,i)$ is denoted by $area_\tau(i)$.
\end{defn}

For a $(k,n)$-tiling $\tau$ the positroid cell $S_{\tau}$ in $Gr_{k,n}(\bF_\bR)$ is constructed in the following way. 
Let $\tau = \{T_1, \dots, T_k\},$ where $T_i = \mathcal{T}_{1,a_i,a_i+1}$ and $a_1 < \dots < a_k$.
We define $S_{\tau}$ as the image of the map $M_{\tau}$ 
\begin{align}
    M_{\tau}: (\bF_\bR^{>0})^{2k} \rightarrow Gr_{k,n}(\bF_\bR),
\end{align}
where $M_{\tau}(x_1, \dots, x_k, y_1, \dots, y_k)$ is the row span of a matrix $M \in Mat_{k,n}(\bF_\bR),$
such that the non-zero entries of $M$ in the $i$-th row are precisely 
\[M_i^1 = 1, M_i^{a_1} = (-1)^{area_{\tau}(i)}x_i, \text{ and } M_i^{a_i +1} = (-1)^{area_{\tau}(i) + k}y_i. \]
\begin{notat}
    We will denote points $(x_1, \dots, x_k, y_1, \dots , y_k)$ by $(\bar{x}, \bar{y})$
\end{notat}
\begin{rmrk}\cite[Theorem 4.17]{PSBW}
    It follows from the explicit construction of $M_\tau$, that for each $I \in \binom{[n]}{k-1}$ such that $I \cap \{1,a_i,a_i+1\} = \varnothing,$ and any $(\bar{x},\bar{y}) \in (\bF_\bR^{>0})^{2k}$  we have 
    \begin{align}\label{eq: kermit_Pluckers}
        M_\tau(\bar{x},\bar{y})^{I \cup \{1\}} =  \frac{1}{x_i} M_\tau(\bar{x},\bar{y})^{I \cup \{a_i\}} = \frac{1}{y_i} M_\tau(\bar{x},\bar{y})^{I \cup \{a_i + 1\}}.
    \end{align}
\end{rmrk}

\begin{thm}\cite[Proposition 4.7, Theorem 4.17]{PSBW}
    \begin{itemize}
        \item The map $M_\tau$ is injective for any $(k,n)$-tiling $\tau$;
        \item For any $I \in \binom{[n]}{k}$, the minor $M_\tau(\bar{x},\bar{y})$ is either zero, or a polynomial with positive coefficients  in $x_1, \dots, x_k, y_1, \dots , y_k$ ;
        \item For any $(k,n)$-tiling $\tau$ the set $S_\tau = M_\tau((\bF_\bR^{>0})^{2k})$ is a positriod cell;
        \item For every $Z \in Mat_{n,k+2}^{>0}(\bF_\bR),$ the collection of cells $\tilde{Z}(S_\tau),$ for all 
        $(k,n)$-tilings, is a triangulation of $\cA_{n,k,2}(Z)$.
    \end{itemize}
\end{thm}
We define the \emph{tropical Kermit positroid cells} $S_\tau^{trop}$ as  valuation images of cells $S_\tau$. 
\begin{thm}\label{thm:trop_kermit}
    The collection of tropical Kermit cells $S_\tau^{trop}$ constitutes a tropical triangulation of $\cA^{trop}_{n,k,2}.$
\end{thm}
The separation of $S_\tau^{trop}$ into $(S_\tau^{trop})^{int}$ and $(S_\tau^{trop})^{bound}$ is given in the next subsection.

\subsection{Weak injectivity of the tropical amplituhedron map on tropical Kermit cells}
In this section we show that $\tilde{Z}$ is weakly injective on $S_\tau^{trop}.$
First, let us construct the tropical analogue of the map $M_{\tau}:$ 

\begin{align*}
    M_{\tau}^{trop}: \bR^{2k} &\rightarrow \tr Gr_{k,n}^{\geq 0}\\
    M_{\tau}^{trop}(X_1, \dots, X_k, Y_1, \dots, Y_k) &= val(M_{\tau}(t^{X_1},\dots,t^{X_k}, t^{Y_1}, \dots, t^{Y_k})).
\end{align*}

\begin{lem}\label{lem:parameterization}
The map $M_\tau^{trop}$ is a homemorphism between $\bR^{2k}$ and the tropical Kermit cell $S_\tau^{trop}.$
\end{lem}
\begin{proof}
This map is clearly continuous. 

Observe that
since the Pl\"ucker coordinates of $M_\tau(\bar{x},\bar{y})$ polynomials with positive coefficients  in $x_1, \dots, x_k, y_1, \dots , y_k$, valuation image of $M_\tau(\bar{x},\bar{y})$ can be expressed directly in terms of $val(x_i), val(y_i)$, and thus the map $M_{\tau}^{trop}$ is surjective onto $val(S_\tau)$.

For injectivity, note that applying valuation to equation \ref{eq: kermit_Pluckers} yields that for each $I \in \binom{[n]}{k-1}$ such that $I \cap \{1,a_i,a_i+1\} = \varnothing$ for any $(\bar{X},\bar{Y}) \in (\bR)^{2k}$  we have 
    \begin{align} \label{eq:kasteleyn-trop-pluck}
        M_\tau^{trop}(\bar{X},\bar{Y})^{I \cup \{1\}} =  M_\tau^{trop}(\bar{X},\bar{Y})^{I \cup \{a_i\}} - X_i = M_\tau^{trop}(\bar{X},\bar{Y})^{I \cup \{a_i + 1\}} - Y_i.
    \end{align}
For a fixed $i \in [k]$, consider $I = \{a_1, a_2, \dots, a_{i-1}, a_{i+1}+1, \dots, a_{k}+1\}$. As $1 < a_1< \dots < a_k,$ we get $I \cap \{1,a_i,a_i+1\} = \varnothing.$ Moreover, since in the matrix $M$ spanning $M_{\tau}(\bar{x},\bar{y})$ the elements $M_j^{a_j}$ and $M_j^{a_j+1}$ are non-zero for all $j \in [k]$, and Pl\"ucker coordinates of $M_\tau(\bar{x}, \bar{y})$ are polynomials in $x_1,\dots,x_k, y_1, \dots, y_k$ with nonnegative coefficients, we see that 
for all $(\bar{x},\bar{y}) \in (\bF_\bR^{>0})^{2k}:$ 
\[M_\tau(\bar{x},\bar{y})^{I \cup \{1\}}, M_\tau(\bar{x},\bar{y})^{I \cup \{a_i\}}, M_\tau(\bar{x},\bar{y})^{I \cup \{a_i + 1\}} \neq 0.\]
Applying valuation to the equation above we obtain that for all $(\bar{X}, \bar{Y}) \in \bR^{2k}:$
\[M^{trop}_\tau(\bar{X},\bar{Y})^{I \cup \{1\}}, M^{trop}_\tau(\bar{X},\bar{Y})^{I \cup \{a_i\}}, M^{trop}_\tau(\bar{X},\bar{Y})^{I \cup \{a_i + 1\}} \neq +\infty.\]
Together with equation \eqref{eq:kasteleyn-trop-pluck} this gives us
\begin{align}
   X_i  =  M_\tau^{trop}(\bar{X},\bar{Y})^{I \cup \{a_i\}} -  M_\tau^{trop}(\bar{X},\bar{Y})^{I \cup \{1\}}, \text{ and }\\
   Y_i = M_\tau^{trop}(\bar{X},\bar{Y})^{I \cup \{a_i + 1\}} - M_\tau^{trop}(\bar{X},\bar{Y})^{I \cup \{1\}},
\end{align}
giving expression for $X_i, Y_i$ in terms of Pl\"ucker coordinates of $M_\tau^{trop}(\bar{X},\bar{Y})$ and thus proving the injectivity of $M_\tau^{trop}.$

Moreover, the previous argument shows that also the inverse map is continuous. So $M_\tau^{trop}$ is indeed a homemorphism.
\end{proof}

Our next step is to construct a map 
\[\varphi_\tau: \cA_{n,k,2}^{trop}(Z) \rightarrow \bR^{2k}, \] 
such that
\[\varphi_\tau\circ \tilde{Z} \circ M_\tau^{trop}    = id\] on the complement of
\begin{align*}
  (\tilde{Z}\circ M_\tau^{trop})^{-1}
  \left(\tilde{Z}(S_\tau^{trop})\cap \left(\bigcup_{i\in[k]}V(\tr \bra Y 1 a_i \ket)\cup   V(\tr \bra Y 1 (a_i+1) \ket)\cup  V(\tr \bra Y a_i (a_i +1)\ket)\right)\right) 
\end{align*}
in $\bR^{2k}.$  
The following diagram is meant to help us keep track of domains and images of $M^{trop}_{\tau}, \tilde{Z}$ and $\varphi$.
\begin{align}
    \begin{CD}
        \bR^{2k} @>M^{trop}_{\tau}>> S_{\tau}^{trop} @>\tilde{Z}>> \cA_{n,k,2}^{trop}(Z) @>\varphi_\tau>> \bR^{2k}.
    \end{CD}
\end{align}

\begin{lem}
    The map $\varphi=\varphi_\tau: \cA_{n,k,2}^{trop}(Z) \rightarrow \bR^{2k}$ defined as
    \begin{align}
        &\forall i \in [k]:\nonumber \\
        &\varphi (Y)_i =  \tr \bra Y 1 (a_i+1) \ket - \tr \bra Y a_i (a_i + 1)\ket,\\
        &\varphi (Y)_{i +k} = \tr \bra Y 1 a_i \ket - \tr \bra Y a_i (a_i + 1)\ket,
    \end{align}
    satisfies the above requirement.
\end{lem}

\begin{proof}
    Consider a point $(\bar{X}, \bar{Y}) \in \bR^{2k}$, such that $\tilde{Z}(M_\tau^{trop}(\bar{X}, \bar{Y}))$ does not belong to the zero loci of tropical twistors $\tr \bra Y 1 a_i \ket$, $\tr \bra Y 1 (a_i +1) \ket$, and $\tr \bra Y a_i (a_i + 1)\ket$ for any $i \in [k]$.
    
    By equation \eqref{eq:uplus_twistors} we have 
    \begin{align}\label{eq:inj_proof_1}
        \tr \bra \tilde{Z}(M_\tau^{trop}(\bar{X}, \bar{Y})) 1 a_i\ket = \min_{I \in \binom{[n]}{k}} (M_\tau^{trop}(\bar{X}, \bar{Y})^I + Z_{I \uplus \{1,a_i\}}^{[k+2]}).
    \end{align}
    Since $\tilde{Z}(M_\tau^{trop}(\bar{X}, \bar{Y})) \notin V(\tr \bra Y 1 a_i \ket),$ by Proposition \ref{not_twistor_zero-no_intersection} this minimum is achieved in some $I$ such that $I \cap \{1,a_i\} = \varnothing$. Since for any $I$ such that $I \cap \{1, a_i, a_i+1\} = \varnothing$ $M^{trop}_\tau(\bar{X},\bar{Y}) = + \infty,$ the minimizing $I$ must contain $a_i +1$. Thus equation \eqref{eq:inj_proof_1} can be rewritten as
    \begin{align}\label{eq:inj_proof_tw1}
        \tr \bra \tilde{Z}(M_\tau^{trop}(\bar{X}, \bar{Y})) 1 a_i\ket = \min_{\substack{I' \in \binom{[n]}{k-1}\\ I' \cap \{1,a_i,a_i+1\} = \varnothing }}(M_\tau^{trop}(\bar{X}, \bar{Y})^{I' \cup \{a_i + 1\}} + Z_{I' \cup \{a_i + 1\} \uplus \{1,a_i\}}^{[k+2]}) = \\
        = \min_{\substack{I' \in \binom{[n]}{k-1}\\ I' \cap \{1,a_i,a_i+1\} = \varnothing} }(M_\tau^{trop}(\bar{X}, \bar{Y})^{I' \cup \{a_i + 1\}} + Z_{I' \cup \{1, a_i, a_i + 1\}} ^{[k+2]})\nonumber
    \end{align}
    In the same way as equation \eqref{eq:inj_proof_tw1}, we obtain 
    \begin{align}\label{eq:inj_proof_tw2}
        \tr \bra \tilde{Z}(M_\tau^{trop}(\bar{X}, \bar{Y})) 1 (a_i + 1)\ket = \min_{\substack{I' \in \binom{[n]}{k-1}\\ I' \cap \{1,a_i,a_i+1\} = \varnothing} }(M_\tau^{trop}(\bar{X}, \bar{Y})^{I' \cup \{a_i\}} + Z_{I' \cup \{1, a_i, a_i + 1\}} ^{[k+2]}),
    \end{align}
    \begin{align}\label{eq:inj_proof_tw3}
        \tr \bra \tilde{Z}(M_\tau^{trop}(\bar{X}, \bar{Y})) a_i (a_i + 1)\ket = \min_{\substack{I' \in \binom{[n]}{k-1}\\ I' \cap \{1,a_i,a_i+1\} = \varnothing} }(M_\tau^{trop}(\bar{X}, \bar{Y})^{I' \cup \{1\}} + Z_{I' \cup \{1, a_i, a_i + 1\}} ^{[k+2]}).
    \end{align}
    Applying valuation to equation \ref{eq: kermit_Pluckers} yields that for each $I \in \binom{[n]}{k-1}$ such that $I \cap \{1,a_i,a_i+1\} = \varnothing$ for any $(\bar{X},\bar{Y}) \in (\bR)^{2k}$  we have 
    \begin{align*}
        M_\tau^{trop}(\bar{X},\bar{Y})^{I \cup \{1\}} =  M_\tau^{trop}(\bar{X},\bar{Y})^{I \cup \{a_i\}} - X_i = M_\tau^{trop}(\bar{X},\bar{Y})^{I \cup \{a_i + 1\}} - Y_i.
    \end{align*}
    This, together with equations \eqref{eq:inj_proof_tw1}-\eqref{eq:inj_proof_tw3} shows that 
    \begin{align*}
        &\tr \bra \tilde{Z}(M_\tau^{trop}(\bar{X}, \bar{Y})) 1 (a_i + 1)\ket - \tr \bra \tilde{Z}(M_\tau^{trop}(\bar{X}, \bar{Y})) a_i (a_i + 1)\ket =  \\ &= \min_{\substack{I' \in \binom{[n]}{k-1}\\ I' \cap \{1,a_i,a_i+1\} = \varnothing} }M_\tau^{trop}(\bar{X}, \bar{Y})^{I' \cup \{a_i\}} - \min_{\substack{I' \in \binom{[n]}{k-1}\\ I' \cap \{1,a_i,a_i+1\} = \varnothing} }M_\tau^{trop}(\bar{X}, \bar{Y})^{I' \cup \{1\}} = X_i, \text{ and }\\
        &\tr \bra \tilde{Z}(M_\tau^{trop}(\bar{X}, \bar{Y})) 1 a_i\ket - \tr \bra \tilde{Z}(M_\tau^{trop}(\bar{X}, \bar{Y})) a_i (a_i + 1)\ket =  \\ &= \min_{\substack{I' \in \binom{[n]}{k-1}\\ I' \cap \{1,a_i,a_i+1\} = \varnothing} }M_\tau^{trop}(\bar{X}, \bar{Y})^{I' \cup \{a_i +1\}} - \min_{\substack{I' \in \binom{[n]}{k-1}\\ I' \cap \{1,a_i,a_i+1\} = \varnothing} }M_\tau^{trop}(\bar{X}, \bar{Y})^{I' \cup \{1\}} = Y_i,
    \end{align*}
    which proves the lemma.
\end{proof}

Denote by $(S_\tau^{trop})^{int}$ the set of $C \in S_\tau^{trop}$, such that $\tilde{Z}(C)$ does not belong to the zero loci of tropical twistors $\tr \bra Y 1 a_i \ket$, $\tr \bra Y 1 (a_i +1) \ket$, and $\tr \bra Y a_i (a_i + 1)\ket$ for any $i \in [k]$. We define $(S_\tau^{trop})^{bound}$ to be the complement $S_\tau^{trop} \setminus (S_\tau^{trop})^{int}$.
\begin{cor}\label{cor:injectivity}
    $(S_\tau^{trop})^{int}$ is open, and the tropical amplituhedron map $\tilde{Z}$ is injective when restricting to it.  
    $\tilde{Z}(S_\tau^{trop})^{bound}$ is contained in the union of zero loci of the tropical twistors $\tr \bra Y 1 a_i \ket$, $\tr \bra Y 1 (a_i +1) \ket$, and $\tr \bra Y a_i (a_i + 1)\ket,$ $i \in [k].$
\end{cor}
\begin{proof}
The complement of the tropical twistors $\tr \bra Y 1 a_i \ket$, $\tr \bra Y 1 (a_i +1) \ket$, and $\tr \bra Y a_i (a_i + 1)\ket,$ $i \in [k]$ is open in $\tr Gr_{k,k+2}.$ Thus the inverse image of this complement under $M_\tau^{trop}\circ \tilde{Z}$ is open in $\bR^{2k}.$ Since $M_\tau^{trop}$ is a homemorphism, by Lemma \ref{lem:parameterization}, also $(S_\tau^{trop})^{int}$ is open.

    Since $\varphi_\tau\circ \tilde{Z} \circ M_\tau^{trop}$ restricted to $(M_\tau^{trop})^{-1}((S_\tau^{trop})^{int})$ is the identity, and $M_\tau^{trop}$ is a bijection, by Lemma \ref{lem:parameterization}, it follows that $\varphi_\tau\circ\tilde{Z}$ is also injective on $(S_\tau^{trop})^{int}$. Thus,  $\tilde{Z}$ must also be injective there. The second statement is just the definition.
\end{proof}

\subsection{Separation of the images of Kermit cells}
In this section we show that cells $S_\tau^{trop}$ satisfy the weak disjointness condition from Definition \ref{def:trop_triangulation}.
\begin{lem}\label{trop_twistor_sign}
    Let $Z$ be a tropically totally positive matrix in $Mat_{n,k+m}(\bR)$, $C \in \tr Gr_{k,n}^{\geq 0}$, and let $C_{\bF_\bR} \in Gr_{k,n}^{\geq 0}$ and $Z_{\bF_\bR}$ be some lifts of $C$ and $Z$.
    Suppose for a set $I = \{i_1, \dots , i_m\} \subset [k+m]$ 
    \[\tilde{Z}(C) \notin V(\tr \bra Y i_1 \dots i_m  \ket)\] and 
    \[ B = \underset{J \in \binom{[n]}{k}}{\arg \min}(C^J + Z_{J \uplus I}^{[k+m]}).\]
    Then the sign $(-1)^{\sigma(B, i_1, \dots, i_m)}$, defined in Remark \ref{rmrk:regular_tw_expansion}, is equal to the sign of $\bra \tilde{Z}_{\bF_\bR}(C_{\bF_\bR}) i_1 \dots i_m\ket.$
\end{lem}

\begin{proof}
    Let us recall equation \eqref{regular_twistors_expansion}:
    \begin{align*}
       \bra \tilde{Z}(C_{\bF_\bR}) i_1 i_2 \dots i_m \ket = \sum_{J \in \binom{[n]}{k}} (-1)^{\sigma(J,\{i_1, \dots , i_m\} )} C_{\bF_\bR}^J   (Z_{\bF_\bR})_{J\cup \{i_1, \dots, i_m\}}^{[k+2]}.
   \end{align*}
   From it we immediately see that the term of $C_{\bF_\bR}^B  (Z_{\bF_\bR})_{B\cup \{i_1, \dots, i_m\}}^{[k+2]}$ the lowest power in $t$ appears with the sign $(-1)^{\sigma(B, i_1, \dots, i_m)}$, meaning that $(-1)^{\sigma(B, i_1, \dots, i_m)}$ is the sign of this twistor.
\end{proof}

\begin{thm}\cite[Theorem 4.14]{PSBW}\label{tw-fixed-sign}
    Let $C_{\bF_\bR} \in S_\tau$. Then for all $i \in [n]$ we have
    \begin{align}
        \text{sgn}\bra \tilde{Z}(C_{\bF_\bR}) 1 i\ket = (-1)^{area_\tau(i)}.    
    \end{align}
\end{thm}
\begin{prop}\label{prop:separation}
    For different $(k,n)$-tilings $\tau$ and $\tau'$ the images of $S_{\tau}^{trop}$ and $S_{\tau'}^{trop}$ under the amplituhedron map are weakly disjoint, meaning that
    \[\dim \tilde{Z}((S_{\tau}^{trop})^{int}) \cap \tilde{Z}((S_{\tau'}^{trop})^{int}) = \varnothing.\]
\end{prop}
Note that the collection of signs $(-1)^{area_\tau(i)}$ uniquely determines $\tau$: $\mathcal{T}_{1,i,i+1} \in \tau$ if and only if $(-1)^{area_\tau(i)} \neq (-1)^{area_\tau(i+1)}$. Therefore, for different $(k,n)$-tilings $\tau$ and $\tau'$ there must exist $i \in [n]$ such that $(-1)^{area_\tau(i)} \neq (-1)^{area_{\tau'}(i)}$. Moreover, we can choose such $i$ on the boundary of a triangle that is contained in at least on of $\tau,~\tau'$. This implies, without loss of generality, that at least one of $\mathcal{T}_{1,i,i+1}$ or $\mathcal{T}_{1,i-1,i}$ belongs to $\tau$.

Let $C \in (S_{\tau}^{trop})^{int}$ and $C' \in (S_{\tau'}^{trop})^{int}$, and fix $i$ such that $(-1)^{area_\tau(i)} \neq (-1)^{area_{\tau'}(i)}$ and $\mathcal{T}_{1,i,i+1}$ or $\mathcal{T}_{1,i-1,i}$ is a triangle in $\tau$.
Suppose 
\begin{align*}
    &B = \underset{J \in \binom{[n]}{k}}{\arg \min}(C^I + Z_{J\uplus \{1,i\}}^{[k+2]}) \text { and }\\
    &B' = \underset{J \in \binom{[n]}{k}}{\arg \min}((C')^I + Z_{J\uplus \{1,i\}}^{[k+2]}).
\end{align*}
Then by Lemma \ref{trop_twistor_sign} and Theorem \ref{tw-fixed-sign} we know 
\begin{align*}
    (-1)^{\sigma(B, 1,i)} = (-1)^{area_\tau(i)} \text{ and }
    (-1)^{\sigma(B', 1,i)} = (-1)^{area_{\tau'}(i)}.
\end{align*}
Thus, $(-1)^{\sigma(B, 1,i)} \neq (-1)^{\sigma(B', 1,i)}$, as $(-1)^{area_\tau(i)} \neq (-1)^{area_{\tau'}(i)}$.
Note that if $B = \{b_1< \dots < b_k\}$ and $q$ is such that $b_q < i < b_{q+1}$, then the sign $(-1)^{\sigma(B, 1,i)}$ coincides with the sign $(-1)^q$. Therefore, if $B' = \{b'_1< \dots < b'_k\}$ and $q'$ is such that $b'_q < i < b'_{q+1}$, then $q \neq q'$. 
By Lemma \ref{diag} we know that 
\begin{align*}
    \tr \bra \tilde{Z}(C) 1 i \ket = \tilde{Z}(C)^{[k+m] \setminus \{1,q+2\}} + Z_{1,i}^{1,q+2}, \text{ and }
    \tr \bra \tilde{Z}(C') 1 i \ket = \tilde{Z}(C')^{[k+m] \setminus \{1,q'+2\}} + Z_{1,i}^{1,q'+2}.
\end{align*}
Let $Y = \tilde{Z}(C)$ and $Y' = \tilde{Z}(C').$
Then from the equations above we have 
\begin{align*}
    \{1,q+2\} \in \underset{S \in \binom{[n]}{2}}{\arg \min}(Y^{[k+m] \setminus S} + Z_{\{1,i\}}^S) \text{ and } \{1,q'+2\} \in \underset{S \in \binom{[n]}{2}}{\arg \min}((Y')^{[k+m] \setminus S} + Z_{\{1,i\}}^S).
\end{align*}
Since we have chosen $i$ such that $\mathcal{T}_{1,i,i+1}$ or $\mathcal{T}_{1,i-1,i}$ is a triangle in $\tau$, by definition of $(S_\tau^{trop})^{int}$ we have $Y \notin V(\tr \bra Y 1 i \ket)$, hence $\underset{S \in \binom{[n]}{2}}{\arg \min}(Y^{[k+m] \setminus S} + Z_{\{1,i\}}^S)$ contains only one element. This shows that 

\[\underset{S \in \binom{[n]}{2}}{\arg \min}(Y^{[k+m] \setminus S} + Z_{\{1,i\}}^S) = \{1,q+2\} \neq \underset{S \in \binom{[n]}{2}}{\arg \min}((Y')^{[k+m] \setminus S} + Z_{\{1,i\}}^S),\]
and thus $Y \neq Y',$
which 
 means 
\[\tilde{Z}((S_{\tau}^{trop})^{int}) \cap \tilde{Z}((S_{\tau'}^{trop})^{int}) = \varnothing.\]

\subsection{Surjectivity}
\begin{prop}\label{surjectivity}
    The closures images of tropical Kermit cells $S_\tau^{trop}$ cover the tropical amplituhedron, namely 
    \[\cA_{n,k,2}^{trop}(Z) = \overline{\tilde{Z}(\bigcup_{(k,n)-tiling~\tau} S_\tau^{trop}).}\]
\end{prop}

\begin{proof}
    Let $Z$ be a tropically totally positive matrix in $Mat_{n, k+2}(\bR)$ and $Z_{\bF_\bR}$ be some lift of $Z$.
    By \cite[Theorem 4.17]{PSBW}(or \cite[Theorem 2.4]{BH}), 
    \begin{align}\label{eq:surj_real}
        \cA_{n,k,2}(Z_{\bF_\bR}) = \overline{\tilde{Z}_{\bF_\bR}(\bigcup_{(k,n)-tiling~\tau} S_\tau)}.
    \end{align}
    
    Applying valuation to equation \eqref{eq:surj_real} by Proposition \ref{amp_trop_well_defined} gives us 
    \[\cA_{n,k,2}^{trop}(Z) = \tilde{Z}(\bigcup_{(k,n)-tiling~\tau} val( \overline{S_\tau})).\]    
    Recall from Definition \ref{def:generalised-puis} that the closure over $\bF_\bR$ is taken with respect to the norm $||f|| = e^{-val(f)}$, and since by \cite[Theorem 11]{Mar} $\bF_\bR$ is complete with respect to this norm, and, therefore, commutes with valuation, giving us 
    \[val(\overline{S_\tau}) = \overline{val(S_\tau)} = \overline{S_\tau^{trop}}.\]
    Since by Proposition \ref{amp_trop_well_defined} \[val(\tilde{Z}_{\bF_\bR}(S_\tau)) = \tilde{Z}(S_\tau^{trop}),\] we have shown that
    \[\cA_{n,k,2}^{trop}(Z) = \overline{\tilde{Z}(\bigcup_{(k,n)-tiling~\tau} S_\tau^{trop}).}\]
\end{proof}

All together, Corollary \ref{cor:injectivity}, Proposition \ref{prop:separation} and Proposition \ref{surjectivity} show that the collection of cells $S_\tau^{trop} = (S_\tau^{trop})^{int}(Z) \sqcup (S_\tau^{trop})^{bound}(Z),$ for all $(k,n)$-tilings, constitutes a tropical triangulation of $\cA_{n,k,2}(Z)$, proving Theorem \ref{thm:trop_kermit}.

\end{document}